\documentclass[aps,prl,amsmath,amssymb,groupedaddress,superscriptaddress,floatfix,twocolumn,showkeys,amsart]{revtex4-1}
\usepackage[utf8]{inputenc}
\usepackage{graphicx} 

\begin{document}

\title{Direct measurement of a spatially varying thermal bath using Brownian motion}

\author{Ravid Shaniv$^*$}
\author{Chris Reetz$^*$}
\author{Cindy A. Regal}

\begin{abstract}
    Micro-mechanical resonator performance is fundamentally limited by the coupling to a thermal environment. The magnitude of this thermodynamical effect is typically considered in accordance with a physical temperature, assumed to be uniform across the resonator's physical span. However, in some circumstances, e.g. quantum optomechanics or interferometric gravitational wave detection, the temperature of the resonator may not be uniform, resulting in the resonator being thermally linked to a spatially varying thermal bath. In this case, the link of a mode of interest to its thermal environment is less straightforward to understand. Here, we engineer a distributed bath on a germane optomechanical platform --- a phononic crystal --- and utilize both highly localized and extended resonator modes to probe the spatially varying bath in entirely different bath regimes. As a result, we observe striking differences in the modes' Brownian motion magnitude. From these measurements we are able to reconstruct the local temperature map across our resonator and measure nanoscale effects on thermal conductivity and radiative cooling. Our work explains some thermal phenomena encountered in optomechanical experiments, e.g. mode-dependent heating due to light absorption. Moreover, our work generalizes the typical figure of merit quantifying the coupling of a resonator mode to its thermal environment from the mechanical dissipation to the overlap between the local dissipation and the local temperature throughout the resonator. This added understanding identifies design principles that can be applied to performance of micro-mechanical resonators.
\end{abstract}

\maketitle

\par Micro-mechanical resonators are at the heart of many technologies, including inertial sensing~\cite{passaro2017gyroscope,krause2012high}, microscopy~\cite{giessibl2003advances,poggio2010force} and bio-sensing \cite{degen2009nanoscale,arlett2011comparative}. Simultaneously, these devices are found to be a paramount tool in fundamental science research, e.g. quantum solid-state experiments~\cite{bleszynski2009persistent}, optomechanical quantum information experiments~\cite{andrews2014bidirectional,wallucks2020quantum,barzanjeh2022optomechanics}, gravity measurements~\cite{schmole2016micromechanical,liu2021gravitational} and dark matter searches \cite{carney2021mechanical}. Because both technological applications and pure scientific studies occur at finite temperature, at all times there is a competition between the desired signal source and the typically undesired excitations of the resonator due to thermal energy, referred to as Brownian motion. It is therefore imperative to understand the behavior of a resonator in contact to a thermal bath.
\par Typically, micro-mechanical resonators are considered as having a uniform temperature identical to the temperature of their immediate environment. In this case the Brownian motion of the $i$ resonator mode, described by coordinate $x_{i}$, is captured by the equipartition theorem,
\begin{equation} \label{eq: Brownian_motion_uniform_T}
    \langle x_{i}^{2} \rangle = \frac{k_{B} T}{m_{i} \omega_{i}^2}.
\end{equation}
Here, $k_{B}$, $m_{i}$ and $\omega_{i}$ are the Boltzmann constant, the mode's effective mass and the mode's natural frequency, respectively, and $T$ is the uniform temperature.
\par In some cases the uniform temperature assumption does not hold, for example in mechanical micro-bolometers \cite{piller2022thermal} or when a mechanical resonator's motion is detected optically. In the latter, when a laser beam is coupled to a resonator mode, the light is partially absorbed locally by the resonator body~\cite{harry2010advanced,riedinger2018remote,mirhosseini2020superconducting,qiu2020laser,peterson2016laser,page2021gravitational}. This can leave the resonator in a thermal non-equilibrium steady state (NESS)~\cite{lax1960fluctuations,komori2018direct}, where temperature is not uniform, and is defined locally. In this case, the temperature $T$ in Eq.~\ref{eq: Brownian_motion_uniform_T} should be replaced with the effective temperature $T^{\left(i\right)}_{\mathrm{eff}}$ of the $i$ mode, defined as:
\begin{equation} \label{eq: Brownian_motion_effective_T}
    T^{\left(i\right)}_{\mathrm{eff}}=\frac{\int_{-\infty}^{\infty}\alpha_{i} T dV}{\int_{-\infty}^{\infty}\alpha_{i}dV}.
\end{equation}
Here both $T$ and $\alpha_{i}$ --- referred to hereafter as the local temperature and dissipation density of the mode $i$ respectively --- are functions of spatial coordinates, the former due to the presence of a heat source and the latter due to the mode displacement function. The denominator $\int_{-\infty}^{\infty}\alpha_{i}dV = \gamma_{i}$ in Eq.~\ref{eq: Brownian_motion_effective_T} is the damping coefficient of the mode $i$. From Eq.~\ref{eq: Brownian_motion_effective_T} it is evident that each resonator mode could have a different effective temperature~\cite{geitner2017low,singh2020detecting}.
\par A key motivation for the study of a resonator's response to spatially varying baths is that Eq.~\ref{eq: Brownian_motion_effective_T} identifies an important figure of merit for the resonator design. Commonly, resonator geometry is engineered to enhance its quality factor, denoted as $Q_{i}$ for the mode $i$~\cite{ghadimi2018elastic,fedorov2020fractal,hoj2021ultra,bereyhi2022perimeter,pluchar2022high,eichenfield2009optomechanical}, as well as maximizing its coupling to a probe. Due to the fact that $Q_{i}$ is defined as the ratio $Q_{i}=\frac{\omega_{i}}{\gamma_{i}}$, many strategies of $Q_{i}$ enhancement aim to increase $\omega_{i}$ and lower $\gamma_{i}$ by geometrical design, while maintaining a high level of mode-probe coupling. However, in any thermally-limited application, the figure of merit for the system's performance would be governed by the quantity $\gamma_{i}T^{\left(i\right)}_{\mathrm{eff}} = \int_{-\infty}^{\infty}\alpha_{i} T dV$, the overlap integral between the local temperature and the dissipation density. This quantity generalizes the thermal decoherence rate of a mode, pertinent to both classical and quantum applications, to $\gamma_{i}n_{\mathrm{th}} = \frac{k_{B}}{\hbar \omega_{i}}\int_{-\infty}^{\infty}\alpha_{i} T dV$, where $n_{th}$ is the steady state average number of thermal phonons. It follows that when a non-uniform temperature profile is expected, optimized performance is obtained by minimizing this overlap rather than minimizing $\gamma_{i}$ alone. Similar analysis is carried out for optimized design of electromagnetic resonators~\cite{clerk2010introduction}.

\par
\begin{figure}[t]
\includegraphics[width=0.48\textwidth]{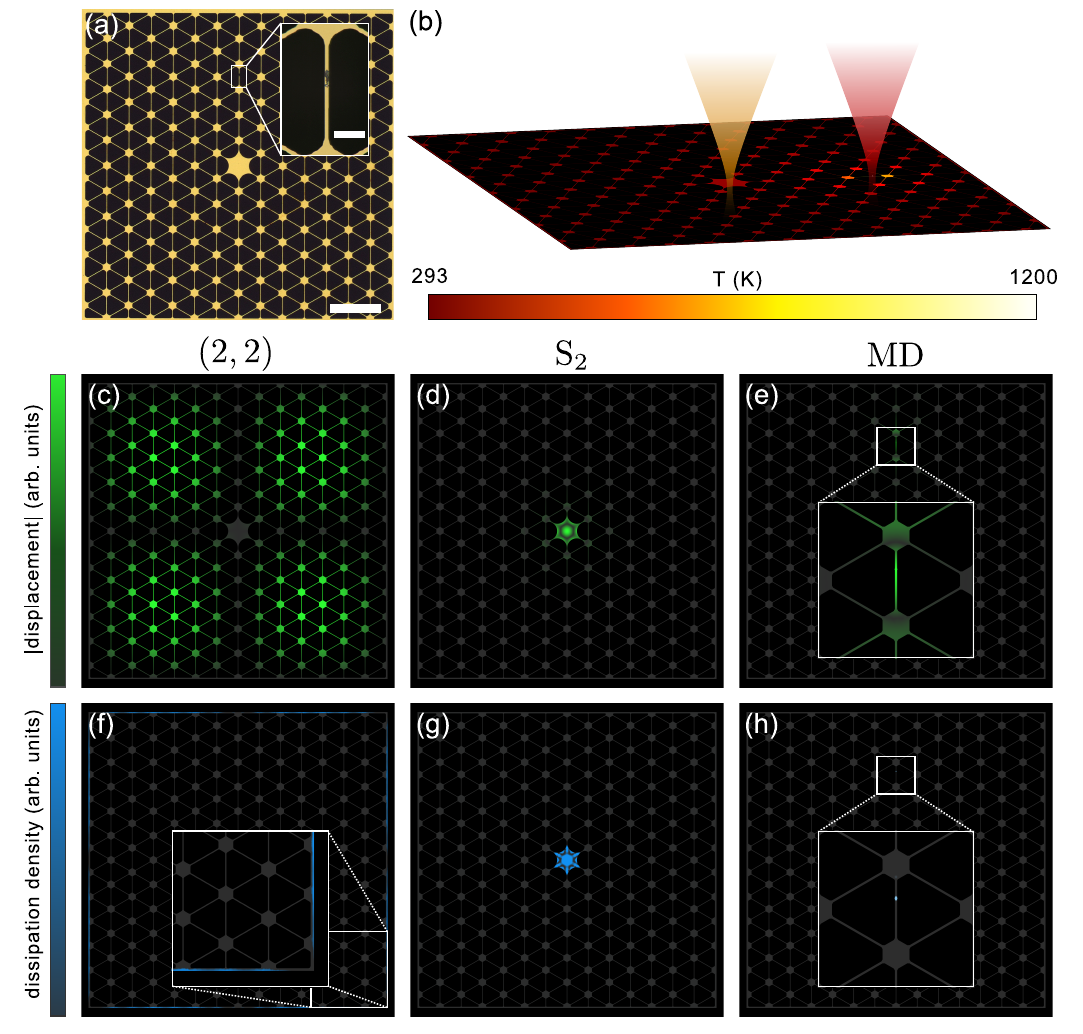}
\caption{(a) Optical microscope image of the device studied in this work (scale bar 200 microns). Inset: zoomed in image of deposited absorber used to generate temperature gradients (scale bar 20 microns). (b) Simulated temperature map observed in this work. A localized heat load was generated by absorbed light from a heating beam (red). The Brownian motion of the device was measured interferometrically with a probe beam (yellow). (c-e) Absolute displacement of the (2,2), $\mathrm{S}_{2}$ and $\mathrm{MD}$ respectively. (f-h) Dissipation density profiles of (2,2), $\mathrm{S}_{2}$ and $\mathrm{MD}$. Inset on (f) shows that the dissipation for the (2,2) mode is concentrated at the edge, while the inset on (h) shows that almost all of the mechanical loss is concentrated in the absorber for the $\mathrm{MD}$.}
\label{Fig: modes_dissipation}
\end{figure}

 Fig.~\ref{Fig: modes_dissipation}(a) shows an optical microscope image of the resonator employed in this work. It is patterned as a phononic crystal (PnC) with a band of forbidden oscillation frequencies - a bandgap \cite{kushwaha1993acoustic,tsaturyan2017ultracoherent}. By placing defects in the crystal pattern (Fig~\ref{Fig: modes_dissipation}(a)), we prepare out-of-bandgap membrane-like modes (Fig.~\ref{Fig: modes_dissipation}(c)) alongside localized modes within the bandgap (Fig.~\ref{Fig: modes_dissipation}(d-e)). These modes have exceptionally different dissipation density (Fig.~\ref{Fig: modes_dissipation}(f-h)). In some cases, the dissipation density differs dramatically from the mode displacement function (Fig.~\ref{Fig: modes_dissipation}(c) and~\ref{Fig: modes_dissipation}(f)), meaning that the effective mode temperature will not be governed by the local temperature where the motion is large, but by where the dissipation is significant.

\par Here, we demonstrate the effect an extreme temperature gradient across a resonator has on the Brownian motion of its different modes. We generate this temperature gradient across a silicon-nitride (SiN) tensioned thin-film resonator by deposition of a localized absorber, heated with laser light (Fig.~\ref{Fig: modes_dissipation}(a) and ~\ref{Fig: modes_dissipation}(b)). Our engineered mode structure enables modes with vastly different effective temperatures to coexist, allowing for direct probing of the temperature across the resonator through the measurement of the different modes' Brownian motion. Furthermore, using Brownian motion as a local temperature probe calibrates emissivity, thermal expansion and thermal conductivity of the oscillator, which have geometry dependent values in nanoscale devices~\cite{zhang2020radiative,leivo1998thermal,vicarelli2022micromechanical,piller2020thermal}. Lastly, we use locally-absorbed heat in order to differentially shift the frequency of localized modes and make two in-bandgap modes hybridize. By hybridizing a pair of localized modes with different temperatures, we increase a mode's effective temperature by in-situ changing its dissipation, rather than its bath temperature.

The PnC pattern was design with have high contrast in order to have a wide bandgap ($>1\: \mathrm{MHz}$) \cite{reetz2019analysis}. The absorber (Stycast 2850ft epoxy) location (Fig.~\ref{Fig: modes_dissipation}(a)) was chosen to satisfy a few requirements. It was deposited on a narrow tether, in order to minimize the rate of heat escaping from it, allowing for greater temperature gradient (Fig.~\ref{Fig: modes_dissipation}(b)) as is common in silicon nitride bolometers\cite{turner2001silicon,vicarelli2022micromechanical}. The specific tether was chosen to minimize the reduction of $\mathrm{Q}$ of resonator modes of interest. Fig.~\ref{Fig: modes_dissipation}(c-e) show the displacement of three of these modes obtained from FEA calculation  - the square-membrane-like $\left(2,2\right)$ mode (Fig.~\ref{Fig: modes_dissipation}(c)) and two in-bandgap modes. One in-bandgap mode has radial symmetry and two radial anti-nodes within the central pad, and denoted hereafter as $\mathrm{S}_{2}$ (Fig.~\ref{Fig: modes_dissipation}(d)). The other in-bandgap mode exists only due to the deposited absorber mass changing the mode structure \cite{hoj2022ultra}, and is denoted as mass defined mode ($\mathrm{MD}$) (Fig.~\ref{Fig: modes_dissipation}(e)). Fig.~\ref{Fig: modes_dissipation}(f-h) show the dissipation density originating from material bending of the corresponding modes. The geometry of the absorber used in FEA simulation was selected to match the observed mechanical resonance frequency of the MD mode. 

\par It is apparent that each mode in Fig.~\ref{Fig: modes_dissipation} experiences its dominant dissipation at a different location. We define  $T_{\mathrm{cp}}$, $T_{\mathrm{ab}}$ and $T_{\mathrm{fr}}$ as the temperature at the center pad, the absorber and the resonator frame respectively. From Eq. \ref{eq: Brownian_motion_effective_T}, if these local temperatures are different, the three aforementioned modes would exhibit different $T_{\mathrm{eff}}$.

\par The resonator was placed in a vacuum chamber, and its motion was measured using a $1064 \mathrm{nm}$ laser and a calibrated  Michelson interferometer, with which we were able to establish the thermal nature of our resonator's motion. A $950 \mathrm{nm}$ wavelength beam is separately aligned for heating the absorber. Both beams' intensities are feedback controlled. Due to the different mode shapes, detection of these modes is done at different locations on the resonator. To reference different measurements with the same heating power, and because absorbed power depends on multiple factors (absorption coefficient, our beam shape and aberrations and our beam alignment), the frequency shift $\Delta f_{1,1}$ of the membrane-like $\left(1,1\right)$ mode is used as a heating power proxy. When the resonator is heated, the SiN thermally expands and its stress lowers, which in turn leads to a frequency drop of its modes, such that larger frequency shift corresponds to more heat being absorbed. $\Delta  f_{1,1}$ was chosen as the $\left(1,1\right)$ mode has detectable motion at any location on the resonator.

\begin{figure}[t]
\includegraphics[width=0.49\textwidth]{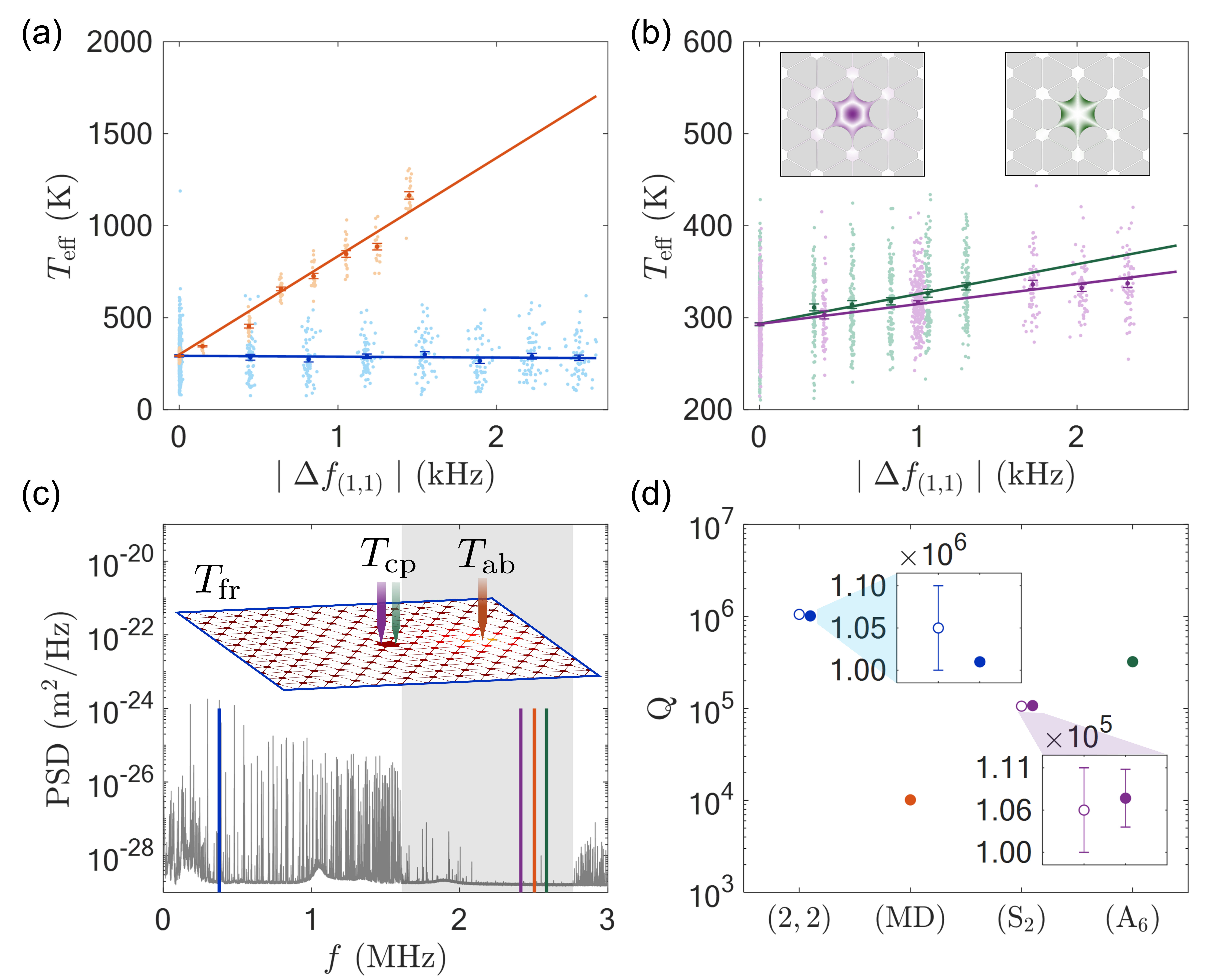}
\caption{$T_{\mathrm{eff}}$ of four modes for different absorber heating. (a) Measured $\tilde{T}^{\left(\mathrm{MD}\right)}_{\mathrm{eff}}$ and $\tilde{T}^{\mathrm{\left(2,2\right)}}_{\mathrm{eff}}$ (orange and blue respectively). (b) Measured $\tilde{T}^{\left(\mathrm{S}_{2}\right)}_{\mathrm{eff}}$ and $\tilde{T}^{\mathrm{A}_{6}}_{\mathrm{eff}}$ (purple and green respectively). The plots show measured (points) and $\Delta f_{1,1}$- binned (full circles with error bars) results. Solid lines are single-parameter linear fits, fixing Brownian motion for $\Delta f_{\left(1,1\right)}=0$ to the lab temperature $T_{\mathrm{lab}}=293\:K$. Insets show the similarity in localization of the absolute displacement of the in-bandgap $\mathrm{S}_{2}$ and $\mathrm{A}_{6}$ modes. (c) Measured power spectral density example of the resonator motion. Colored lines mark the frequencies of the four modes, and shaded region marks identifies the frequency bandgap. The inset schematically shows the location of temperature each mode probes (color code is similar to (a) and (b)). (d) Measured $Q$ values of the four modes (color code is similar to (a) and (b)). Empty (Full) circles show measured $Q$ before (after) absorber deposition. Error bars not shown means they are smaller than the point marker size.}
\label{Fig:heating_results}
\end{figure}

\par Results of heating experiments are shown in Fig.~\ref{Fig:heating_results}. By measuring the Brownian motion $\langle x_{i}^{2}\rangle$ of mode $i$ and it's corresponding mode angular frequency $\omega_{i}$ for different laser heating powers - and therefore different $\Delta f_{1,1}$ - we can define the random variable $\tilde{T}^{\left(i\right)}_{\mathrm{eff}}$ representing the measured mode's effective temperature:

\begin{equation} \label{eq: Effective_temperature_definition}
    \tilde{T}^{\left(i\right)}_{\mathrm{eff}} = \frac{\langle x_{i}^{2}\rangle \omega^{2}_{i}}{\langle\langle x_{i,0}^{2}\rangle  \omega^{2}_{i,0} \rangle_{\mathrm{all}}}T_{\mathrm{lab}},
\end{equation}
derived using Eq.~\ref{eq: Brownian_motion_uniform_T}. Here, $\langle \cdot \rangle$ denotes average over a single-shot $0.3 \: \mathrm{s}$ time interval, chosen for technical reasons and $\langle \cdot \rangle_{\mathrm{all}}$ denotes a full average over all the no-heating data, which was taken with large statistics and is assumed to have negligible variance. $\langle x_{i,0}^{2}\rangle$ and $\omega_{i,0}$ are the measured displacement power and angular frequency of the mode with no absorber heating, respectively, and $T_{\mathrm{lab}}$ is the lab temperature measured with a thermometer to be $293 \mathrm{K}$.
\par Fig.~\ref{Fig:heating_results}(a) shows measured $\tilde{T}^{\left(\mathrm{MD}\right)}_{\mathrm{eff}}$ and $\tilde{T}^{\left(2,2\right)}_{\mathrm{eff}}$ for different $\left(1,1\right)$ mode frequency shift. In striking contrast to the mode displacement, the dissipation density of the $\left(2,2\right)$ mode is localized at the resonator boundary (Fig.~\ref{Fig: modes_dissipation}(f)) which is characteristic to an out-of-bandgap low frequency membrane-like mode \cite{tsaturyan2017ultracoherent,reetz2019analysis}. As a result, $\tilde{T}^{\left(2,2\right)}_{\mathrm{eff}}$ probes the temperature $T_{\mathrm{fr}}=T_{\mathrm{lab}}$, and indeed $\tilde{T}^{\left(2,2\right)}_{\mathrm{eff}}$ does not deviate from the lab temperature at various heating powers. In contrast, the $\mathrm{MD}$ dissipation density is localized at the heating point (Fig.~\ref{Fig: modes_dissipation}(h)), which exhibits the highest temperature $T_{\mathrm{ab}}$ upon heating. Therefore, a sizeable difference is observed between $\tilde{T}^{\left(\mathrm{MD}\right)}_{\mathrm{eff}}$ and $\tilde{T}^{\left(2,2\right)}_{\mathrm{eff}}$ as heating power increases. Fig.~\ref{Fig:heating_results}(b) shows the result of a similar heating experiment measured on the $\mathrm{S}_{2}$ mode. This mode dissipation density is localized at the middle pad (Fig.~\ref{Fig: modes_dissipation}(g)), probing $T_{\mathrm{cp}}$, and therefore $\tilde{T}^{\left(\mathrm{S}_{2}\right)}_{\mathrm{eff}}$ is greater than the lab temperature, but lower than $\tilde{T}^{\left(\mathrm{MD}\right)}_{\mathrm{eff}}$. 
\par In order to verify that the heating of the $\mathrm{S}_{2}$ mode is due to the temperature at the center pad and not due to the small dissipation density at the hot absorber location, we measured the heating of a second mode, having six azimuthal anti-nodes at the central pad, which we denoted as $\mathrm{A}_{6}$. This mode's dissipation density is confined to the center pad similarly to the $\mathrm{S}_{2}$ mode, but its residual dissipation density at the absorber position is significantly different. From Fig.~\ref{Fig:heating_results}(b) it can be seen that both $\tilde{T}^{\left(\mathrm{S}_{2}\right)}_{\mathrm{eff}}$ and $\tilde{T}^{\left(\mathrm{A}_{6}\right)}_{\mathrm{eff}}$ rise at a similar rate with respect to $\Delta f_{1,1}$ (within statistical error). Fig.~\ref{Fig:heating_results}(c) is an example spectrum of resonator modes, exhibiting a bandgap (shaded grey area). The frequencies of the different modes are marked (colored lines), supporting the localization of the in-bandgap modes. As a secondary test, we compared $Q$ measurements of the $\mathrm{S}_{2}$ and the $\left(2,2\right)$ modes before and after the absorber deposition. The $Q$ value did not change within the measurement error bars, which agrees with the dissipation density contribution of the absorber in the $\mathrm{S}_{2}$ and $\left(2,2\right)$ modes being minor. We therefore substantiated that $\tilde{T}^{\left(\mathrm{S}_{2}\right)}_{\mathrm{eff}}$ probes the local temperature at the center pad.

\par
\begin{figure}[t]
\centering
\includegraphics[width=0.45\textwidth]{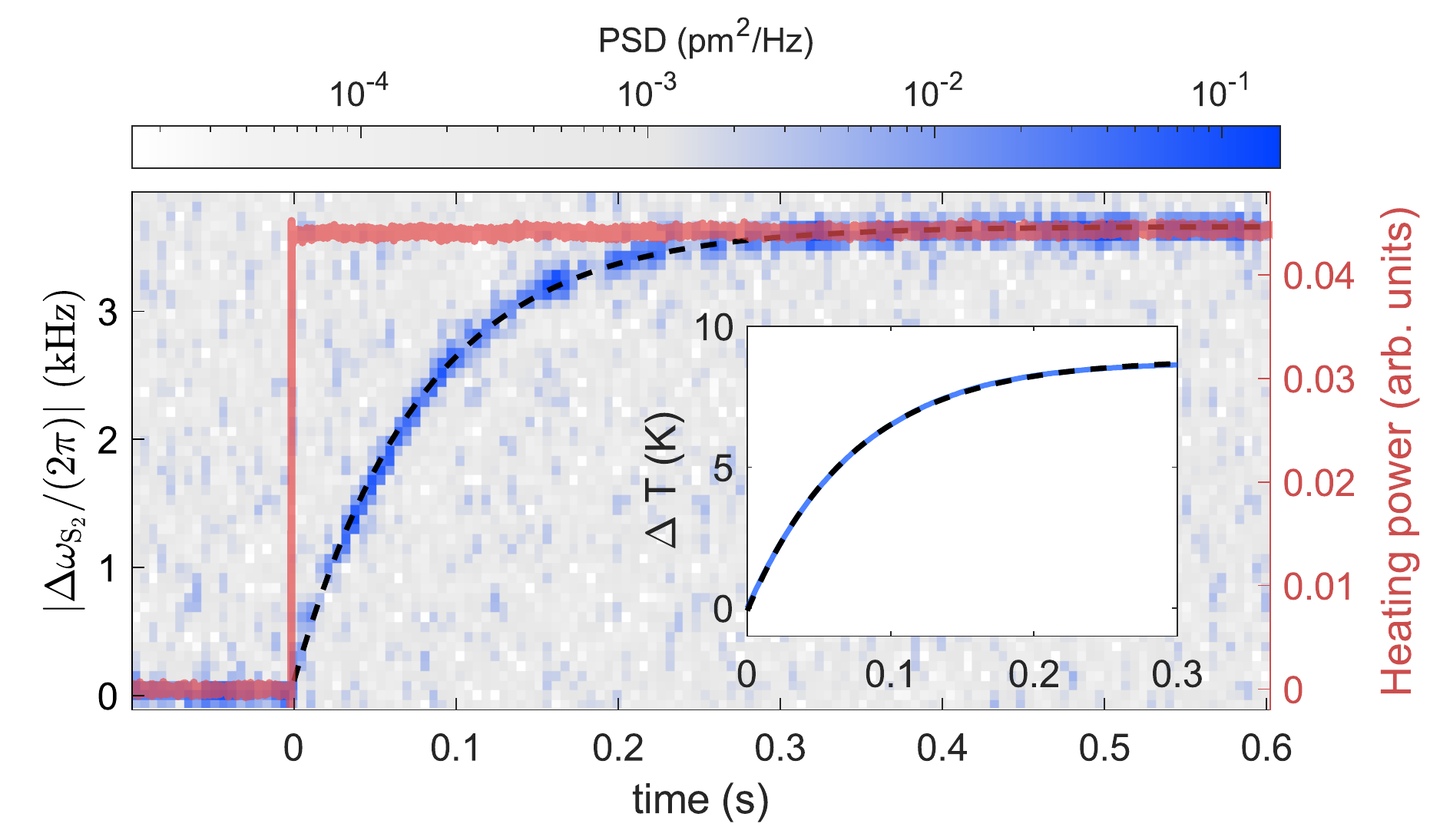}
\caption{Determination of thermalization time scale $\tau_{\mathrm{th}}$ from step function response of $\Delta\omega_{\mathrm{S}_{2}}$. Left axis: Spectrogram of thermomechanical motion. The dashed black line shows an exponential fit to the peak power of each time bin of the spectrogram. The fitted rate of the frequency shift $1/\tau_{\mathrm{th}}$ is 12.5 Hz. Right axis: monitor of the heating laser power during this experiment. Inset: The simulated time dependence of the temperature increase at the defect pad center (blue line). Example exponential fit to this curve (black dashed line).}
\label{fig:freqTimescale}
\end{figure}

 Next, we turn to using these results and obtain the spatial temperature distribution across our resonator. A full description of the steady state temperature map of the heat equation requires knowledge of the thermal conductivity $k_{c}$, emissivity $\varepsilon$ and heat load $P_{\mathrm{heat}}$. Measuring both $\tilde{T}^{\left(\mathrm{S}_{2}\right)}_{\mathrm{eff}}$ and $\tilde{T}^{\left(\mathrm{MD}\right)}_{\mathrm{eff}}$ provides the physical temperatures $T_{\mathrm{cp}}$ and $T_{\mathrm{ab}}$ respectively. These two independent measurements are insufficient to fully constrain the three parameters of the heat equation. A measurement of the thermalization time scale $\tau_{\mathrm{th}}$ provides additional knowledge about $k_{c}$ and $\varepsilon$ independent of $P_{\mathrm{heat}}$. Fig.~\ref{fig:freqTimescale} shows the $\mathrm{S}_{2}$ mode frequency shift $\Delta \omega_{\mathrm{S}_{2}}$ when the absorber is subjected to a step heat load. The frequencies quasistatically follow the instantaneous temperature profile of the device, and thus the timescale of the frequency change is equal to $\tau_{\mathrm{th}}$. Fitting the result to an FEA simulation of the time dependent temperature profile allows for a necessary constraint $k_{c}(\varepsilon)$ needed to explain the value of $1/\tau_{\mathrm{th}}$ = 12.5 Hz (inset Fig.~\ref{fig:freqTimescale}). This knowledge in addition to the aforementioned two independent local temperature measurements generates the temperature map across our resonator for various heating powers. Once the temperature maps are known, FEA simulations of the normal mode frequencies can be matched to experiment for a determination of the coefficient of thermal expansion as well (detailed calculation of the temperature map along with an example of measurement-based FEA temperature map are given in Appendix A). 

\par The effective temperature of the modes chosen for our analysis thus far was essentially affected by a single physical temperature - the local temperature at the location of their confined dissipation density (Fig.~\ref{Fig: modes_dissipation} and Fig.~\ref{Fig:heating_results}(b). As a simplest further study, we examined the case in which a single mode is influenced by two different local temperatures, which was made possible owing to the non-uniform local temperature.

\par When a mechanical resonator is held at a uniform temperature, a change in that temperature would lead to identical relative frequency shift for all the modes, meaning $\frac{\Delta f_{i}}{f_{i}}$ is identical for any mode $i$, where $\Delta f_{i}$ is the change of frequency $f_{i}$ due to the temperature change. In this case, two modes would never cross in frequency. However, non-uniform heating generates non-uniform thermal stress change, such that the fractional frequency shift differs between modes \cite{jockel2011spectroscopy,st2019swept,sadeghi2020thermal}, and they may cross in frequency. Indeed, in our device, for some absorber heating the $\mathrm{MD}$ and $\mathrm{S}_{2}$ fractional shifts satisfy $\frac{\Delta f_{\mathrm{MD}}}{f_{\mathrm{MD}}}\approx 13\: \frac{\Delta f_{\mathrm{S}_{2}}}{f_{\mathrm{S}_{2}}}$.

\par
\begin{figure}[b]
\includegraphics[width=0.49\textwidth]{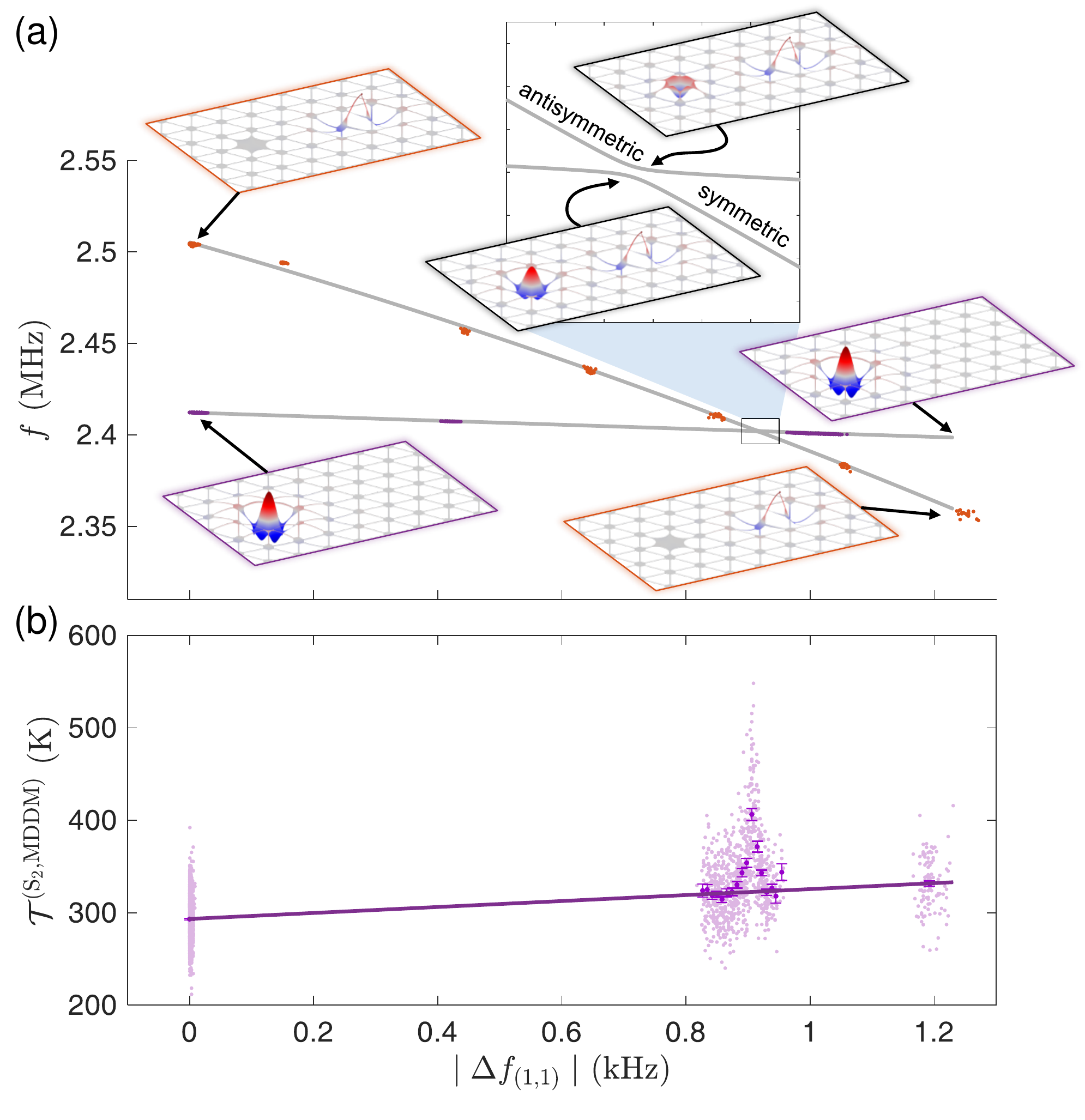}
\caption{Hybridization of $\mathrm{S}_{2}$ and $\mathrm{MD}$ modes. (a) Measured frequencies of $\mathrm{S}_{2}$ and $\mathrm{MD}$ modes with respect to $\Delta f_{1,1}$ (orange and purple points). Gray lines mark a theoretical hybridization frequency curves obtained from fits to the measured modes' frequencies and FEA prediction for their coupling strength. Mode-hybridization shape from FEA are shown along the frequency curves. Inset shows a zoomed-in plot around the full hybridization point. (b) $\tilde{\mathcal{T}}^{\left(\mathrm{S}_{2},\mathrm{MD}\right)}_{\mathrm{eff}}$ as a function of $\Delta f_{1,1}$, Measured (purple points) and $\Delta f_{1,1}$-binned (purple full circles with error bars). The line is a single parameter fit, disregarding all points around $\delta f_{1,1}\approx 0.9\:\mathrm{kHz}$. (a) and (b) share the horizontal axis.}
\label{Fig: Hybridization_results}
\end{figure}

 This allows us to examine the hybridization of these two in-bandgap modes, as a mode coupled to two thermal baths. As shown above, these two modes dissipation densities are restricted to different regions of the resonator, held at different local temperature when the absorber is heated. The frequencies of the $\mathrm{S}_{2}$ and $\mathrm{MD}$ modes for different heating power are shown in Fig.~\ref{Fig: Hybridization_results}(a). The $\mathrm{S}_{2}$ and the $\mathrm{MD}$ frequencies were fitted and extrapolated with linear and quadratic polynomials, respectively. The quadratic polynomial was used because of a mode at $2.465\:\mathrm{MHz}$, with which the $\mathrm{MD}$ mode hybridized, making the $\mathrm{MD}$ frequency shift non linear. The hybridization gap of $\approx500 \:\mathrm{Hz}$ (on par with similar related studies~\cite{catalini2020soft}) between the symmetric and antisymmetric branches (inset in Fig.~\ref{Fig: Hybridization_results}(a)) and the associated mode shapes were estimated from an FEA calculation.

\par As a result of the hybridization, each of two modes in the pair changes its effective mass. Inference of effective temperature from displacement power should, in principle, take this change into account. This requires precise modeling and stable heating, which is challenging. To circumvent these requirements, we examine the quantity

\begin{equation} \label{definition_of_y_square}
     y_{\left(\mathrm{S}_{2},\mathrm{MD}\right)}^{2} \equiv  x_{\left(\mathrm{S}_{2}\right)}^{2} + x_{\left(\mathrm{MD}\right)}^{2},
\end{equation}

which is defined with respect to the center pad motion. In words, $y_{\left(\mathrm{S}_{2},\mathrm{MD}\right)}^{2}$ is the total displacement power of the resonator middle pad at both modes' frequencies. Furthermore, we can define the parameter

\begin{equation} \label{eq: two_modes_effective_temperature}
    \tilde{\mathcal{T}}^{\left(\mathrm{S}_{2},\mathrm{MD}\right)}_{\mathrm{eff}} \equiv
    \frac{\langle y_{\left(\mathrm{S}_{2},\mathrm{MD}\right)}^{2}\rangle \omega^{2}_{\mathrm{S}_{2}}}{\langle\langle y_{\left(\mathrm{S}_{2},\mathrm{MD}\right),{0}}^{2}\rangle \omega^{2}_{\mathrm{S}_{2},0}\rangle_{\mathrm{all}}}T_{\mathrm{lab}},
\end{equation}

which has units of temperature and is defined similarly to the definition in Eq.~\ref{eq: Effective_temperature_definition}. Here, $\omega_{\mathrm{S}_{2}}$ is the extrapolated angular frequency of the $\mathrm{S}_{2}$ mode without hybridization. As it turns out, this temperature does not depend on the effective mass changes due to the mode coupling, yet can still carry information about the local temperatures of the two non-hybridized modes. A detailed derivation of $\tilde{\mathcal{T}}^{\left(\mathrm{S}_{2},\mathrm{MD}\right)}_{\mathrm{eff}}$ is given in Appendix B. In general, it depends on the dissipation, the effective mass, the frequency, and the coupling between the two modes, as well as the local bath temperature of each mode, namely $T_{\mathrm{cp}}$ and $T_{\mathrm{ab}}$. It is, however, constructive to write the expected value for $\tilde{\mathcal{T}}^{\left(\mathrm{S}_{2},\mathrm{MD}\right)}_{\mathrm{eff}}$ in two specific limits:

\begin{equation} \label{eq: effective_two_modes_temperature_limits}
\tilde{\mathcal{T}}^{\left(\mathrm{S}_{2},\mathrm{MD}\right)}_{\mathrm{eff}} = 
     \begin{cases}
        T &\quad T_{\mathrm{cp}}=T_{\mathrm{ab}}=T \\
        \frac{T_{\mathrm{cp}}\gamma_{\mathrm{S}_{2}}+T_{\mathrm{ab}}\gamma_{\mathrm{MD}}}{\gamma_{\mathrm{S}_{2}}+\gamma_{\mathrm{MD}}} &\quad\omega_{\mathrm{S}_{2}}=\omega_{\mathrm{MD}} \\
     \end{cases}.
\end{equation}

Fig.~\ref{Fig: Hybridization_results}(b) shows $\tilde{\mathcal{T}}^{\left(\mathrm{S}_{2},\mathrm{MD}\right)}_{\mathrm{eff}}$ as a function of $\Delta f_{1,1}$. The line is a single parameter fit, taking only the measurements around $\Delta f_{1,1}=0$ and $\Delta_{1,1}\approx 1.2\: \mathrm{kHz}$, which are far detuned from the point of hybridization. This line, which agrees with the linear fit in Fig.~\ref{Fig:heating_results}(b), stands for $T_{\mathrm{cp}}$ inferred from $y_{\left(\mathrm{S}_{2},\mathrm{MD}\right)}^{2}$. According to Eq.~\ref{eq: effective_two_modes_temperature_limits}, a deviation from this line necessarily means that $T_{\mathrm{cp}}\ne T_{\mathrm{sb}}$, meaning that a single hybridized mode is affected by two different thermal baths. The height of the peak around $\Delta f_{1,1}$ is lower than expected value of $750 \mathrm{K}$ according to the second limit in Eq.~\ref{eq: effective_two_modes_temperature_limits} (Appendix B). We believe this is due to heating power fluctuations, originating primarily from motion of the heating beam and temporal changes absorber heating (see Appendix D).

\par To conclude, in this work we designed and implemented a controlled local temperature measurement of a SiN resonator under localized heating, by measuring the Brownian motion of a set of localized normal modes. We experimentally demonstrated that in the presence of a non-uniform temperature profile, different modes might have exceedingly different effective temperatures, depending on the spatial overlap between the local temperature and the dissipation density of a mode. This demonstration was done both by measuring the Brownian motion of different modes and by in-situ hybridization of two modes with different local baths. We compared between  a membrane-like mode, having effective temperature identical to the resonator's environment with the price of higher dissipation at its edge, to a localized mode, potentially designed for lower dissipation, with the price of an effective temperature equal to the local --- typically hotter --- temperature at the mode's confined location. These represent two extreme scenarios of susceptibility to uneven heating. A resonator's optimal performance would be achieved when its design minimizes the overlap between the dissipation density and the local temperature profile. 

$*$ R.S and C.R contributed equally to this work.

\begin{acknowledgments}
We thank Maxwell Urmey, Albert Schliesser, Sofia Brown, and Sanjay Kumar Keshava for helpful discussions, and Nicholas Frattini for careful reading of the manuscript. This work was supported by funding from NSF Grant No. PHYS 1734006, Cottrell FRED Award from the Research Corporation for Science Advancement under grant 27321, and the Baur-SPIE Endowed Professor at JILA. R.~S.~acknowledges support from the Israel Council for Higher Education.
\end{acknowledgments}

\section{Appendix A: Calibration of temperature maps}

In this work, spatially varying temperature profiles were generated by heating a localized absorber on the device. Due to the relatively large temperatures generated in this work ($\sim$1000 K), effects such as thermal radiative cooling needed to be considered to model an arbitrary temperature map. To model such a map, the heat equation with source terms is considered:

\begin{equation}
    \label{eq:heatEquation}
    \rho C_{p} \dfrac{\partial T}{\partial t} + \nabla \cdot \mathbf{q} = \dot{\mathcal{Q}}(x,y,z)
\end{equation}

Here $\rho$ is the material density, $C_{p}$ is the specific heat capacity at constant pressure, $\mathbf{q}$ is the heat flux and $\dot{\mathcal{Q}}(x,y,z)$ is the heat source due to laser light absorption. When considering effects of radiative cooling (or heating), then $\mathbf{q}$ is:

\begin{equation}
    \label{eq:heatFlux}
    \mathbf{q} = -k_{c}\nabla T - \mathbf{n} \sigma \varepsilon(x,y,z) (T_{env}^{4}-T^{4})
\end{equation}

where $k_{c}$ is the coefficient of thermal conductivity, $\mathbf{n}$ is the surface normal vector, $\varepsilon(x,y,z)$ is the surface emissivity and $\sigma$ is the Stefan Boltzmann constant. Note that in this model $\varepsilon(x,y,z) = \varepsilon_{Si_{3}N_{4}}$ on the surface of the material, and is equal to $0$ otherwise. 

\begin{figure}[h!]
\centering
\includegraphics[width=0.38\textwidth]{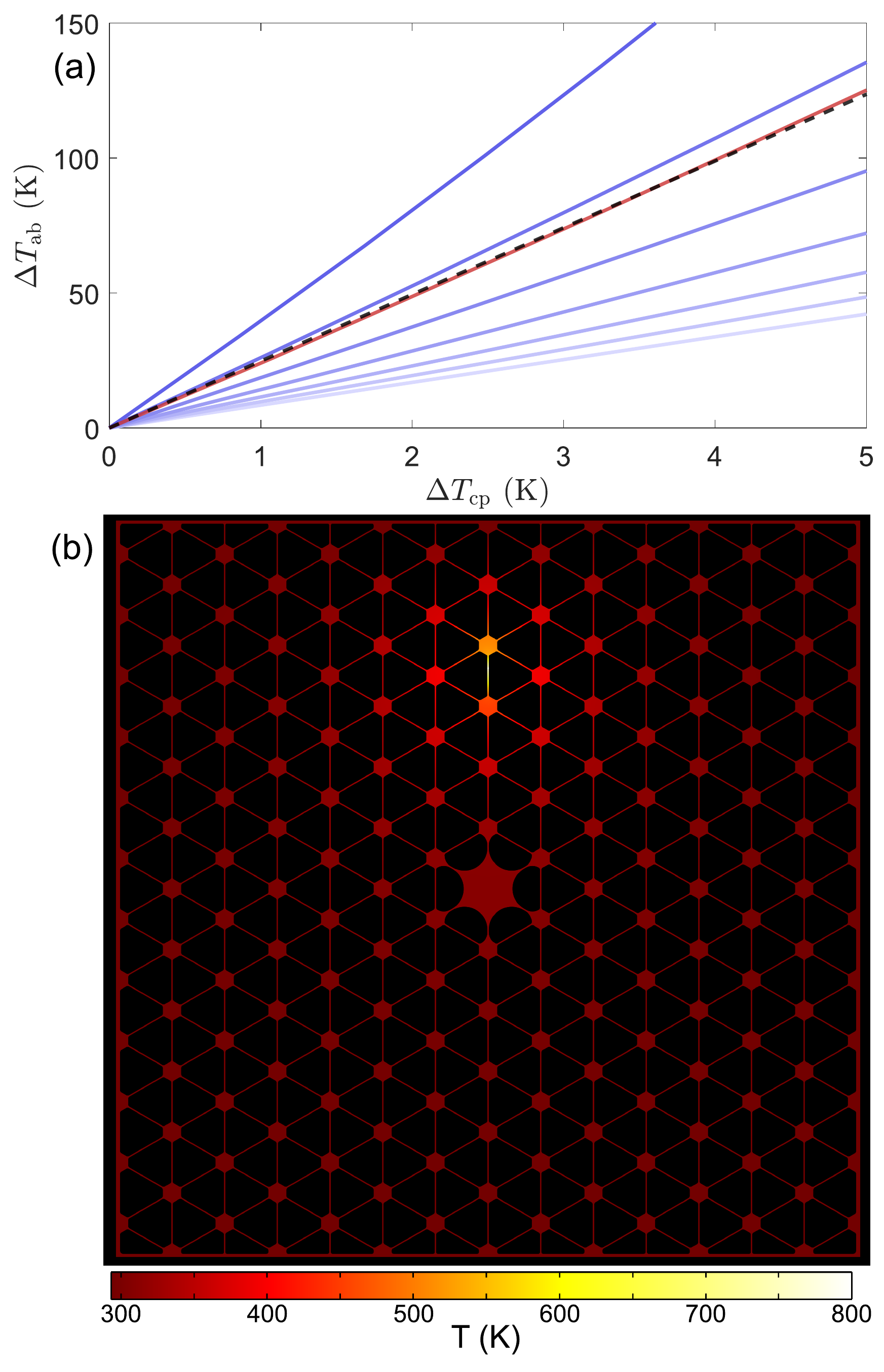}
\caption{(a) Temperature increase of the absorber
 versus the temperature increase of the central pad for increasing heating power. Purple lines correspond to finite element analysis (FEA) simulations where the values of $\varepsilon$ and $k_{c}$ were consistent with the observed $\tau_{\mathrm{th}}$. Light to dark purple indicate the trend when varying $\varepsilon$ from 0 to 0.15. The black dashed shows the inferred local temperature increase from the measured modal temperatures, while the red line shows the best fit FEA simulation. (b) Example simulated temperature map assuming a heat load of 10 $\mathrm{\mu W}$.}
\label{fig:MDvsS2}
\end{figure}

Equations \ref{eq:heatEquation} and \ref{eq:heatFlux} depend on several material properties that were not known precisely for the device studied in this work, namely $k_{c}$ and $\varepsilon$. Geometric dependencies --- and therefore spatial dependencies --- of these parameters have been observed and modelled. Notably there are discrepancies between the bulk values and those observed in thin film systems and narrow constrictions\cite{zhang2020radiative,leivo1998thermal}. However, for all simulations to follow, these values are taken to be uniform across the device, and therefore can be thought of as effective parameters for this specific geometry. Additionally, the precise value of $\dot{\mathcal{Q}}$ is also considered unknown due to possible imperfect knowledge about the absorber and heating laser beam.

Estimates of the parameters are obtained by comparing FEA simulation to measurements. In all FEA simulations performed in this work, the absorber was modelled as a 3 micron diameter sphere, neglecting finer details of the absorber. This geometry matched the observed modal frequency of the $\mathrm{MD}$ mode with simulation. The heat load was considered to be spatially uniform across the entire volume of the sphere: $\dot{\mathcal{Q}} = P_{\mathrm{heat}}/V_{\mathrm{abs}}$. Here $P_{\mathrm{heat}}$ is the total absorbed power from the laser and $V_{\mathrm{abs}}$ is the volume of the absorber. Finally, the boundary condition at the interface between the membrane edge and substrate was held constant at $T_{\mathrm{lab}}$.
\begin{table}[b]
    \centering
    \begin{tabular}{c c c}
        \hline
        thermal parameter& & value \\
        \hline
        $k_{c}$ & & 2.2 W/(m$\cdot$K)\\ 
        $\varepsilon_{\mathrm{Si_{4}N_{3}}}$& & 0.12 \\ 
        $C_{p,\mathrm{Si_{4}N_{3}}}$& & 700 J/($\mathrm{kg \cdot K}$) \\ 
        $\alpha_{\mathrm{\mathrm{th}}}$& & $\mathrm{1.9\times 10^{-6}}$ ($\mathrm{K}^{-1}$)\\
        \hline
        mechanical parameter & & value\\
        \hline
        $\rho_{\mathrm{Si_{4}N_{3}}}$& & 3100 kg/(m$^{3}$)\\
        $E_{\mathrm{Si_{4}N_{3}}}$ & & 250 GPa \\
        $\sigma_{\mathrm{Si_{4}N_{3}}}$ & & 1.05 GPa\\
        $\nu_{\mathrm{Si_{4}N_{3}}}$ & & 0.23
    \end{tabular}
    \caption{Thermal and mechanical parameters of $\mathrm{Si_{4}N_{3}}$ used to produce temperature maps used in this work. $k_{c}$ is the thermal conductivity, $\varepsilon_{\mathrm{Si_{4}N_{3}}}$ is the emissivity, $C_{p,\mathrm{Si_{4}N_{3}}}$ is the heat capacity at constant pressure and $\alpha_{\mathrm{\mathrm{th}}}$ is the coefficient of thermal expansion. $\rho_{\mathrm{Si_{4}N_{3}}}$ is the bulk density, $E_{\mathrm{Si_{4}N_{3}}}$ is the Young's modulus, $\sigma_{\mathrm{Si_{4}N_{3}}}$ is the tensile stress and $\nu_{\mathrm{Si_{4}N_{3}}}$ is the Poisson ratio of the stoichiometric LPCVD silicon nitride used in this work.}
    \label{tab:SiNThermalParams}
\end{table}

One salient experiment is to study the frequency shift in response to a step heating of the device. If the thermalization timescale $\tau_{\mathrm{th}}$ is much shorter than the timescale of stress redistribution then the instantaneous mechanical frequency will adiabatically follow the time-dependent temperature profile. Thus the timescale of the frequency shift $\tau_{\omega_{m}}$ should be equal to $\tau_{\mathrm{th}}$. For a tensioned membrane device, the timescale of stress redistribution is on the order of $L/c \approx 1 \mu s$, where $L$ is the device size and $c$ is the speed of sound. As can be seen in Fig.~\ref{fig:freqTimescale}, the observed timescale is on the order of 100 $ms$, and therefore we can infer that $\tau_{\mathrm{th}} = \tau_{\omega_{m}}$. Finite element simulations of the time-dependent temperature profile can be performed over a large parameter range of $\varepsilon$ and $k_{c}$, an example of which is presented in the inset of Fig.~\ref{fig:freqTimescale}. It was found that there was a one-dimensional manifold of $(\varepsilon,k_{c})$ pairs that fit the observed value of $\tau_{\mathrm{th}}$.

\begin{figure}
    \centering
    \includegraphics[width=0.45\textwidth]{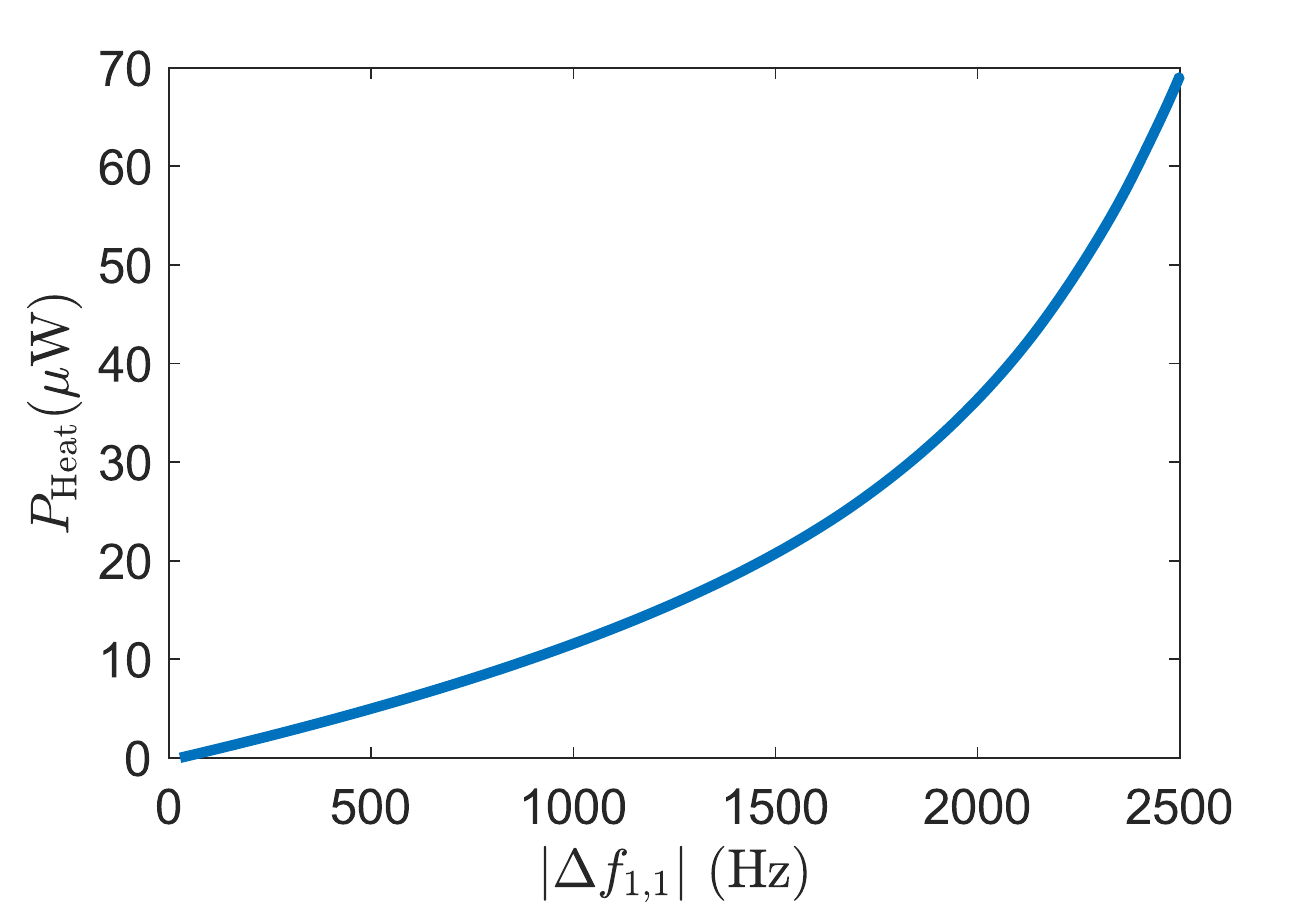}
    \caption{Simulated heating power $P_{\mathrm{heat}}$ with respect to the simulated frequency shift of the (1,1) mode $\Delta f_{\mathrm{1,1}}$. This dependence can be used as a calibration for heating power as a function of observed frequnecy shift. The change in functional form at larger frequency shifts is due to the transition between the heat transfer being conduction dominated to radiative cooling dominated.}
    \label{fig:freq_to_heat_cal}
\end{figure}

Another bound on these parameters comes from the relative modal temperatures of the $\mathrm{S}_{2}$ and $\mathrm{MD}$  modes, which probe the local temperatures at the center pad and the absorber respectively. As can be seen in Fig.~\ref{fig:MDvsS2}, the functional form of $T_{\mathrm{ab}}(T_{\mathrm{cp}})$ depends strongly on the emissivity of the silicon nitride; higher values of $\varepsilon$ produce large temperature gradients between the absorber and defect pad due to radiative cooling near the absorbing region. For this study, observed modal temperatures and simulated physical temperatures can be directly compared since temperature variations over the central pad and absorber regions are relatively small. Matching the slope of the observed $T^{\left(\mathrm{MD}\right)}_{\mathrm{eff}} (T^{\left(\mathrm{S}_{2}\right)}_{\mathrm{eff}})$ to FEA simulations allows for the determination of both $\varepsilon_{\mathrm{Si_{4}N_{3}}}$ and $k_{c}$. Notably, the measured parameters in Tab.~\ref{tab:SiNThermalParams} closely match those measured or calculated in other works~\cite{zhang2020radiative,vicarelli2022micromechanical}. Once the temperature map can be determined, the frequency shift of the (1,1) mode as a function of heating power can also be matched to simulation for a calibration of the coefficient of thermal expansion of $\mathrm{Si_{4}N_{3}}$(Fig.~\ref{fig:freq_to_heat_cal}).

\section{Appendix B: coupled oscillators subject to spatial varying baths} 
\label{Appendix: Coupled_oscillators_temperature}
\subsection{Equivalence of coupled continuum normal modes to coupled point mass oscillators}
Here we will consider the case of two coupled modes of a tensioned mechanical device. It has been shown that the full elastic dynamics associated with these two modes can be reduced to a set of coupled differential equations connecting the modal amplitudes of the modes in question~\cite{catalini2020soft}:

\begin{equation}
    \label{eq:coupledEOM}
    \begin{split}
        K_{11}u_{1}+K_{12}u_{2}&=M_{11}\ddot{u}_{1}+M_{12}\ddot{u}_{2}\\
        K_{21}u_{1}+K_{22}u_{2}&=M_{21}\ddot{u}_{1}+M_{22}\ddot{u}_{2}
    \end{split}
\end{equation}

where the entries for the modal mass matrix $M_{ij}$ and the modal stiffness matrix $K_{ij}$ are given as:

\begin{equation}
M_{nm} = \rho\langle\phi_{n}\lvert\phi_{m}\rangle
\end{equation}
\begin{equation}
K_{nm} = \sigma \langle\nabla \phi_{n}\lvert\nabla \phi_{m}\rangle
\end{equation}

where the $\langle\cdot|\cdot\rangle$ corresponds to the volumetric overlap integral of the quantities in question. 

Eq.~\ref{eq:coupledEOM} can be rearranged to the more familiar form:

\begin{equation}
    \label{eq:Hookes}
    \mathcal{M} \ddot{u} + \mathcal{K} u = 0
\end{equation}

where $\mathcal{M}$ is given as:
\begin{equation}
\label{eq:reducedmassmatrix}
\begin{pmatrix}
    M_{11}-\frac{M_{12}M_{21}}{M_{22}}& 0\\
    0&M_{22}-\frac{M_{12}M_{21}}{M_{11}}
\end{pmatrix}
\end{equation}

and $\mathcal{K}$ is:
\begin{equation}
\label{eq:stiffnessmatrix}
\begin{pmatrix}
    K_{11}-\frac{K_{21}M_{12}}{M_{22}}& K_{12}-\frac{K_{22}M_{12}}{M_{22}}\\
    K_{21}-\frac{K_{11}M_{21}}{M_{11}}&K_{22}-\frac{K_{12}M_{21}}{M_{11}}
\end{pmatrix}
\end{equation}

$u$ is a column vector with componenets $u_{1}$ and $u_{2}$. For modes that are well localized and spatially separated, the overlap integrals  are small: $M_{11},M_{22} \gg M_{12}=M_{21} , K_{11},K_{22} \gg K_{12} = K_{21}$. Also, since we are interested in the behavior where hybridization may occur -- and thus the frequencies of the two modes are nearly degenerate -- it follows that $K_{11}/M_{11}\approx K_{22}/M_{22}$.

Taking the leading order terms in the coupling, it follows that we can write $\mathcal{M}$ and $\mathcal{K}$ as:

\begin{equation}
\label{eq:simplereducedMatrixM}
\mathcal{M} 
=
\begin{pmatrix}
    M_{11}& 0\\
    0&M_{22}
\end{pmatrix}\equiv
\begin{pmatrix}
    m_{1}& 0\\
    0&m_{2}
\end{pmatrix}
\end{equation}

\begin{equation}
\label{eq:simplereducedMatrixK}
\mathcal{K}=
\begin{pmatrix}
    K_{11}& K_{12}-\frac{K_{22}M_{12}}{M_{22}}\\
    K_{21}-\frac{K_{11}M_{21}}{M_{11}}&K_{22}
\end{pmatrix}
\equiv
\begin{pmatrix}
    k_{1}& -\kappa\\
    -\kappa & k_{2}
\end{pmatrix}
\end{equation}

An inspection of Eqs.~\ref{eq:simplereducedMatrixM} and \ref{eq:simplereducedMatrixK} shows that two coupled continuum mechanical modes with small overlap can be reduced to two coupled simple harmonic oscillators.

\subsection{Coupled damped mechanical harmonic oscillators}

In this section, we consider the effects of damping on the hybridization of two point mass coupled oscillators. To model this system, the equations of motion are defined as:

\begin{equation}
\mathcal{M} \ddot{u} + \mathcal{C} \dot{u} + \mathcal{K} u = 0.
\end{equation}

Here we define the mass matrix $M$, the damping matrix $C$ and the spring matrix $K$ as:

\begin{equation}
    \begin{split}
        \mathcal{M} & = 
        \begin{pmatrix}
            m_{1} & 0 \\
            0 & m_{2}
        \end{pmatrix} \\
        \mathcal{C} & = 
        \begin{pmatrix}
            m_{1} \gamma_{1} & 0 \\
            0 & m_{2} \gamma_{2} 
        \end{pmatrix} \\
        \mathcal{K} & = 
        \begin{pmatrix}
            m_{1} \omega_{1}^{2} &  -\sqrt{m_{1} m_{2}} \, g^{2} \\
             -\sqrt{m_{1} m_{2}} \, g^{2} & m_{2} \omega_{2}^{2}
        \end{pmatrix} \\
    \end{split}
\end{equation}

Note that the convention for coupling terms in $K$ is selected such that the normal mode splitting at zero detuning is $g^{2}/\omega_{0}$ for all values of $m_{1}$ and $m_{2}$ in the undamped case. 

To calculate the normal modes, one can assume that $u(t) = u_{0} e^{\lambda t}$. In this case, the equations of motion can be rephrased as a polynomial in $\lambda$ with matrix coefficients:

\begin{equation}
\label{eq:polyEigenvalueProblem}
(\mathcal{M} \lambda^2 + \mathcal{C} \lambda + \mathcal{K})u_{0} = 0.
\end{equation}

Much like an eigenvalue problem, this equation has nontrivial solutions for both $\lambda$ and $u_{0}$ if $\lambda$ is a root of the following polynomial:

\begin{equation}
\label{eq:characterisiticPoly}
\det(\mathcal{M} \lambda^2 + \mathcal{C} \lambda + \mathcal{K}) = 0
\end{equation}

In general, there are four solutions for $\lambda$ for the above equation, coming in two complex conjugate pairs. Physically, the imaginary part of $\lambda$ corresponds to the frequency of each mode, while the real part corresponds to the energy decay rate of the mode in question. Although there is an analytical expression for each $\lambda$, its form is rather involved and will not be presented in this work. Once the roots of Eq.~\ref{eq:characterisiticPoly} are known, then inserting each root into Eq.~\ref{eq:polyEigenvalueProblem} produces a system of linear equations whose null space contains the normal mode corresponding to the eigenvalue in question. The two complex conjugate pair solutions for $\lambda$ produce a complex conjugate pair of normal modes up to a scale factor that without loss of generality can be neglected. 

\begin{figure}[t]
    \centering
    \includegraphics[width=0.48\textwidth]{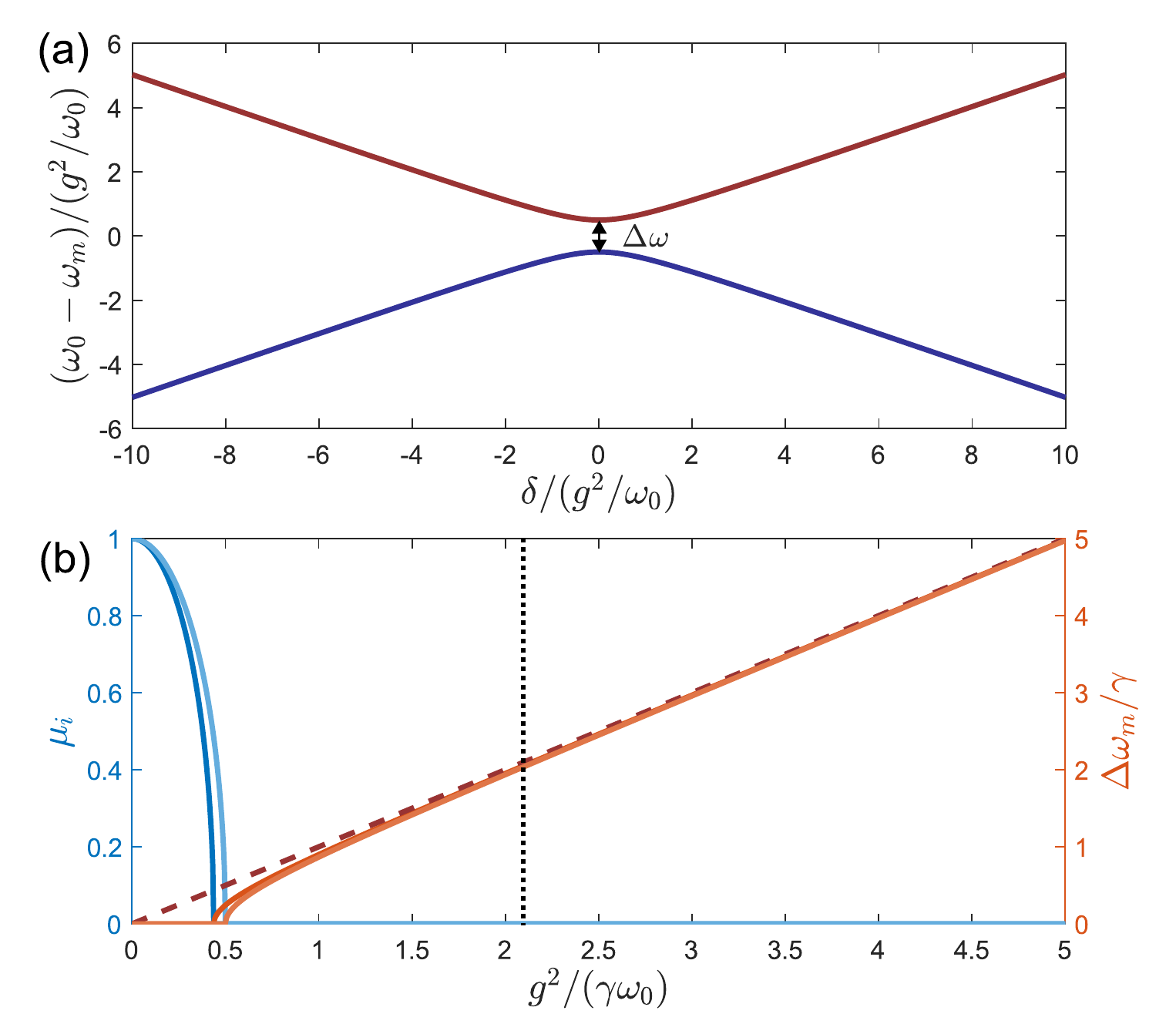}
    \caption{(a) An example of an avoided crossing when $g \gg \gamma \omega_{0}/2$. The red (blue) line corresponds to the normal mode frequency associated with the antisymmetric (symmetric) normal mode. (b) Left axis: mixing factor $\mu_{i}$ as a function of coupling strength $g$. Right axis: normal mode splitting $\Delta \omega_{m}$ as a function of coupling $g$ in the case when  $\gamma_{2} = \gamma \gg \gamma_{1}$. Dark (light) lines correspond to the behaviour in the case when $\gamma_{1} = 0$ ($\gamma_{1} = \gamma_{2}/8$). The dotted red line indicates the asymptotic value of the splitting expected when $\gamma_{1} = \gamma{2} = 0$. The latter case corresponds to the parameters explored experimentally. In both cases, there is a critical value of $g$, below which an avoided crossing does not occur. In the case of no avoided crossing, it is evident that the degree of hybridization also decreases since $\mu_{i} > 0$ in this regime. The vertical dotted line corresponds to the value of $g$ between the $\mathrm{S}_{2}$ mode and the mass defined defect mode (MD) explored experimentally in this work.}
    \label{fig:criticalSplitting}
\end{figure}

We draw attention to the parameter regime where $\gamma_{2} \rightarrow \gamma \gg \gamma_{1} \rightarrow 0$. Here it is evident that there is a critical value of $g = g_{c} = \gamma \omega_{0}/2$ below which there is no frequency splitting (see Fig.~\ref{fig:criticalSplitting}). It is evident that this absence of splitting is correlated to a lowered degree of mode hybridization. This can be quantified by examining the mixing factor $\mu_{i}$ as a function of the mode detuning $\delta$:

\begin{equation}
    \mu_{i} \equiv \min{(||u_{1}^{(i)}(\delta)|^2 - |u_{2}^{(i)}(\delta)|^2 |)}.
\end{equation}

$\mu_{i}$ ranges between 0 and 1, and $u_{j}^{(i)}$ is motional amplitude of mass $j$ in normal mode $i$. When $\mu_{i} = 0$, this means that there is a detuning at which the mode is fully hybridized (equal participation of masses $m_{1}$ and $m_{2}$, while when $\mu_{i} = 1$, this means that the mode has participation of only a single mass. It can be seen in Fig.~\ref{fig:criticalSplitting} that $\mu_{i} > 0$ only occurs when the avoided crossing disappears. 

An important point about this model is that the value of $g$ can be calculated from the undamped behavior of the system. Notably, the coupling strength between the $\mathrm{S}_{2}$ and $\mathrm{MD}$ modes considered in this work can then be taken directly from FEM simulations that neglect damping. The simulated coupling strength between these two modes corresponds to behavior that is close to the undamped regime of mode coupling.

In this section, it is assumed that the normal modes of the system are known. At the outset, the masses, damping rates, and temperatures (expressed as $m_{i}$, $\gamma_{i}$, and $T_{i}$ respectively) of the local modes are known. The goal will be to derive the analogous properties ($M_{i}$, $\Gamma_{i}$ and $T_{\mathrm{eff}}^{(i)}$) of the normal modes.

The damping rates of the normal modes can be calculated from energetic arguments. Notably, it can be interpreted as the energy lost per oscillation times the oscillation rate:

\begin{equation}
    \label{eq:energyGamma}
    \Gamma = \omega_{m} \frac{\Delta W}{2\pi W}
\end{equation}

where $\Delta W$ is the energy lost per oscillation, and $W$ is the energy stored in the oscillator. In this coupled mode model, $\Delta W$ can be calculated as the sum of the work done by damping forces on each point mass per cycle. For harmonic motion, $W$ can be calculated to be twice the average kinetic energy over a single oscillation period. Therefore in the case of two coupled point masses, the expression for the damping rate of mode $i$ is then:

\begin{equation}
    \label{eq:pointMassGamma}
    \begin{split}
    \Gamma_{i} = & \gamma_{1} m_{1} \frac{\rvert u_{1}^{(i)}\lvert^{2}}{(\rvert u_{1}^{(i)}\lvert ^{2} m_{1}+\rvert u_{2}^{(i)}\lvert ^{2} m_{2})} +\\ & \gamma_2 m_{2} \frac{\rvert u_{2}^{(i)}\lvert ^{2}}{(\rvert u_{1}^{(i)}\lvert ^{2} m_{1}+\lvert u_{2}^{(i)}\rvert ^{2} m_{2})}
    \end{split}
\end{equation}

To calculate the effective temperature of the mode in this model, we begin with the general case of a continuous oscillator subject to a spatially varying thermal bath:

\begin{equation}
    T_{\mathrm{eff}} = \frac{\int \alpha T dV}{\int \alpha dV} = \frac{\int \alpha T dV}{\Gamma}
\end{equation}

where $T$ is the local physical temperature of the mechanical structure, and $\alpha$ is the dissipation density of the mode. The analogous formula for the coupled point oscillator model would replace integrals over volume with summations over the contributions from each mass:

\begin{equation}
    T_{\mathrm{eff}} = \frac{\int \alpha T dV}{\Gamma} \rightarrow \frac{\sum_{i=1}^{2} \tilde{\alpha}_{i} T_{i}}{\Gamma}
\end{equation}

Here we have neglected the mode indices on $T_{\mathrm{eff}}$ and $\alpha$ for clarity. An inspection of Eq.~\ref{eq:pointMassGamma} readily identifies an expression for $\tilde{\alpha}_{i}$:

\begin{equation}
    \tilde{\alpha}_{i} = \frac{\gamma_{i}m_{i}|u_{i}|^{2}}{|u_{1}|^{2}m_{1}+|u_{2}|^{2}m_{2}}
\end{equation}

Therefore, the effective temperature for normal mode $i$ is:

\begin{equation}
    T_{\mathrm{eff}}^{(i)} = \frac{T_1 \gamma_{1} m_{1} |u_{1}^{(i)}|^2 + T_2 \gamma_{2} m_{2} |u_{2}^{(i)}|^2}{\gamma_{1} m_{1} |u_{1}^{(i)}|^2 +  \gamma_2 m_{2} |u_{2}^{(i)}|^2}
\end{equation}

Note that in this calculation, normalization conventions for the mode shape vector components $u_{i,j}$ have no effect on the result. Furthermore, this result relies only on the hybridized mode shape and thus is valid for all regimes of the hybridization process. 

To infer the Brownian motion from the effective temperature, the effective mass of each mode must be considered. We note that the effective mass depends not only on the mode being probed but also on the probe location. In the point mass case, there are two probe locations --- one for each mass --- and thus we can define the effective mass of mode $i$ observed at mass $j$ to be:

\begin{equation}
M_{eff,j}^{(i)} = m_1 \frac{|u_{1}^{(i)}|^2}{|u_{j}^{(i)}|^2} + m_2 \frac{|u_{2}^{(i)}|^2}{|u_{j}^{(i)}|^2}
\end{equation}

Finally, the observed Brownian motion can be expressed from the equipartition theorem:

\begin{equation}
    \langle (x_{j}^{(i)})^{2} \rangle = \frac{k_{B} T_{\mathrm{eff}}^{(i)}}{M_{eff,j}^{(i)} \Omega_{i}^{2}}
\end{equation}

When experimentally probing the effects of hybridization on the modal temperatures, a salient quantity to consider is $y_{i}$:

\begin{equation}
\langle y_{i}^{2} \rangle = \langle (x_{i}^{(1)})^{2} \rangle + \langle (x_{i}^{(2)})^{2} \rangle
\end{equation}

From the above formalism, it can be shown that if the normal modes are computed in the undamped limit:

\begin{equation}
\begin{split}
&\langle y_{i}^{2} \rangle = \frac
{k_{B} \mathcal{T}_{\mathrm{eff}}^{(i)}}
{m_{i} \omega_{i}^{2}}\\
&\mathcal{T}_{\mathrm{eff}}^{(i)} = T_{i}
\frac{(\frac{T_{j}}{T_{i}}-1)\frac{\omega_{i}^{2}}{\omega_{j}^{2}}+\frac{\gamma_{j}T_{j}}{\gamma_{i}T_{i}}+2+\frac{\gamma_{i}}{\gamma_{j}}+{(\frac{\omega_{i}^{2}-\omega_{j}^{2}}{g^{2}})}^{2}}
{\bigl(1-\frac{g^{4}}{\omega_{i}^{2}\omega_{j}^{2}}\bigl)\Bigl((\frac{\gamma_{i}}{\gamma_{j}}+2+\frac{\gamma_{j}}{\gamma_{i}})+{(\frac{\omega_{i}^{2}-\omega_{j}^{2}}{g^{2}})}^{2}\Bigl)}
\end{split}
\end{equation}

This expression for $\langle y_{i}^{2} \rangle $ has the property that it depends only on $m_{i}$, the mass of the non-hybridized mode $i$.
Another notable property is revealed when considering the case that $T_{i} = T_{j} = T$:

\begin{equation}
\begin{split}
\langle y_{i}^{2} \rangle &= \frac
{k_{B} T}
{m_{i} \omega_{i}^{2}}{\biggl(1-\frac{g^{4}}{\omega_{i}^{2}\omega_{j}^{2}}\biggl)}^{-1} \\
&= \frac
{k_{B} T}
{m_{i} \omega_{i}^{2}}\biggl(1+\mathcal{O}(\frac{g^{4}}{\omega_{i}^{2}\omega_{j}^{2}})\biggl)
\end{split}
\end{equation}

For this work, the simulated minimal normal mode splitting is 500 Hz, therefore $\frac{g^{4}}{\omega_{i}^{2}\omega_{j}^{2}} \approx 10^{-8}$. Hence, a measurement of $\langle y_{i}^{2} \rangle$ has no discernible dependence on detuning when bath temperatures are equal, regardless of any other parameter mismatch between the modes in question. Therefore, any change in $\langle y_{i}^{2} \rangle$ necessarily arises from a mismatch in thermal bath temperatures. When the two oscillators are degenerate ($\omega_{i} = \omega_{j}$), the expression for $\langle y_{i}^{2} \rangle$ reduces to the simple form:

\begin{equation}
\langle y_{i}^{2} \rangle = \frac
{k_{B}}
{m_{i} \omega_{i}^{2}}\biggl( \frac{T_{i}\gamma_{i}+T_{j}\gamma_{j}}{\gamma_{i}+\gamma_{j}} + \mathcal{O}(\frac{g^{4}}{\omega_{i}^{4}})\biggl)
\end{equation}

This expression of the quantity $\langle y_{i}^{2} \rangle$ corresponds to a single local oscillator subject to two different baths. The inferred temperature of this local oscillator would be:

\begin{equation}
\mathcal{T}_{\mathrm{eff}}^{(i)} \approx \frac{T_{i}\gamma_{i}+T_{j}\gamma_{j}}{\gamma_{i}+\gamma_{j}} 
\end{equation}

The above expression reduces to Eq.~\ref{eq: effective_two_modes_temperature_limits} for $i \rightarrow \mathrm{S}_{2}$ and $j \rightarrow \mathrm{MD}$. 
\section{Appendix C: Additional measurements and overview of mechanical modes}

\begin{figure}[t]
\centering
\includegraphics[width=0.45\textwidth]{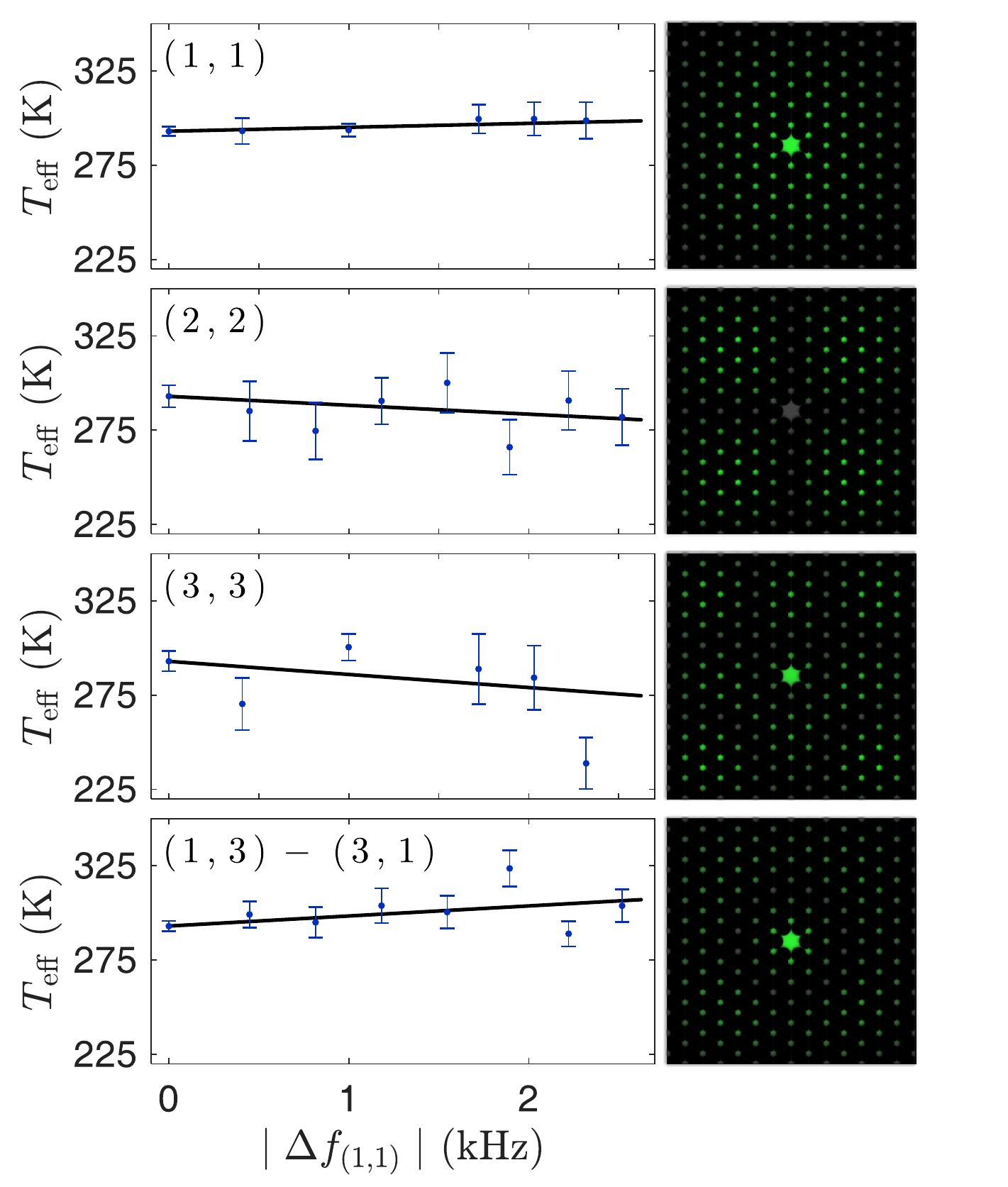}
\caption{Effective temperature of membrane-like modes along with the absolute displacement of each mode. $\Delta f_{1,1}$- binned (full blue circles with error bars) $\tilde{T}^{\left(\mathrm{i}\right)}_{\mathrm{eff}}$ for $i = \left(1,1\right), \:\left(2,2\right),\:\left(3,3\right)$ and $\left(1,3\right)-\left(3,1\right)$ modes.}
\label{fig: membrane_modes_heating}
\end{figure}

In this work modal temperatures of various mechanical modes were measured when subject to varying temperature maps. In addition to the modes presented in the main text, Fig.~\ref{fig: membrane_modes_heating} shows $\tilde{T}^{\left(\mathrm{i}\right)}_{\mathrm{eff}}$ for $i = \left(1,1\right), \:\left(2,2\right),\:\left(3,3\right)$ and $\left(1,3\right)-\left(3,1\right)$, which are membrane-like modes, at various heating powers (referenced by the frequency shift of $\Delta f_{1,1}$). The $(1,3) - (3,1)$ label refers to an antisymmetric hybridization of the (1,3) and (3,1) membrane modes that arises due to the particular patterning of the device. It is evident that these modes do not exhibit an observable increase in their effective temperature with measurement uncertainty. In addition, Tab.~\ref{tab:summary} provides general parameters regarding the modes measured in this work, for completion. It can be understood from Tab.~\ref{tab:summary} that the (1,1) and $(1,3) - (3,1)$ membrane modes have reduced quality factors even before deposition. One expects that the material loss limited quality factor of lower order membrane modes to be on the order for $2\times 10^{6}$ for such a PnC device, and therefore these diminished quality factors can be attributed to losses beyond the suspended silicon nitride structure, commonly referred to as radiation loss. Because heating of the absorber does not increase the substrate temperature appreciably, we do not expect radiation loss limited modes to exhibit elevated effective temperature~\cite{singh2020detecting}.

The (3,3) membrane mode experienced an increase in net dissipation after deposition. However, the lack of heating observed on the (3,3) mode is indicative that this increase in dissipation can also be attributed to radiation loss: if this increase in dissipation is attributed to the addition of the absorber, modal heating would be expected. 

\begin{table}[h!]
    \centering
    \begin{tabular}{c c c c c}
         mode & $\omega_{m}/(2\pi)$ & $m_{eff,opt}$ & $Q_{\mathrm{bare}}$ & $Q_{\mathrm{loaded}}$ \\
         \hline
         (1,1) & 181 kHz & 15.5 ng & $\mathrm{21} \times \mathrm{10^{3}} $ & $\mathrm{32} \times \mathrm{10^{3}} $ \\
         (2,2) & 376 kHz & 17.0 ng & $\mathrm{1.05} \times \mathrm{10^{6}}$ & $\mathrm{1.07} \times \mathrm{10^{6}}$ \\
         (1,3) - (3,1) & 389 kHz & 7.3 ng & $\mathrm{65} \times \mathrm{10^{3}}$ & $\mathrm{29} \times \mathrm{10^{3}}$ \\
         (3,3) & 542 kHz & 14.9 ng & $\mathrm{2.07} \times \mathrm{10^{6}}$ & $\mathrm{1.96} \times \mathrm{10^{6}}$ \\
         $\mathrm{S_{2}}$ & 2.41 MHz & 0.41 ng & $\mathrm{106} \times \mathrm{10^{3}}$ & $\mathrm{107} \times \mathrm{10^{3}}$ \\
         $\mathrm{MD}$ & 2.45 MHz & 0.1 ng & $ - $ & $\mathrm{12.5} \times \mathrm{10^{3}}$ \\
         $\mathrm{A_{6}}$ & 2.58 MHz & 0.3 ng & $ - $ & $\mathrm{322} \times \mathrm{10^{3}}$ \\
         
    \end{tabular}
    \caption{Table of modal parameters for various modes studied in this work. $\omega_{m}$ is the angular frequency of each mode. $m_{eff,opt}$ is the effective mass of each mode when probing at the point of maximum mechanical amplitude. $Q_{\mathrm{bare}}$ ($Q_{\mathrm{loaded}}$) is the mechanical quality factor measured before (after) deposition of the absorber.}
    \label{tab:summary}
\end{table}

\section{Appendix D: Heating power fluctuations}

\begin{figure}[t]
    \centering
    \includegraphics[width=0.48\textwidth]{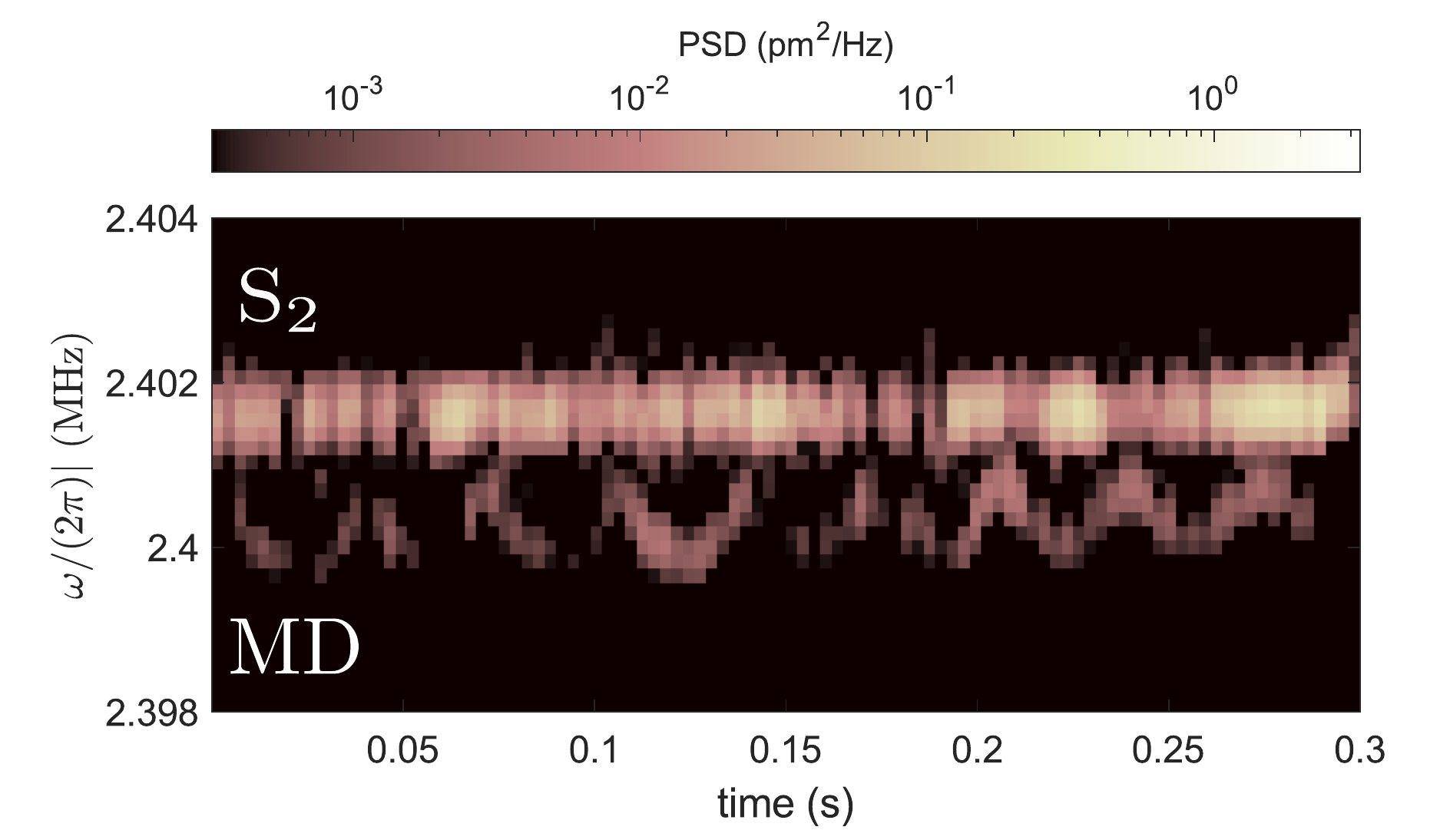}
    \caption{Spectrogram of thermomechanical motion acquired while probing on the central defect pad while heating such that the $\mathrm{S}_{2}$ and MD modes are closely hybridized. The MD mode has relatively large frequency fluctuations due to its relatively large fractional frequency shift.}
    \label{fig:freqFluctuations}
\end{figure}

Throughout this work laser heating was used to generated temperature gradients across the membrane device that are assumed to not vary in time. Under this assumption, the state of the system can be assumed to be in a non equilibrium steady state. Experimental precautions were taken in order to mitigate noise associated with intensity fluctuations of both the probe and heating beams, namely intensity feedback control of both beams. However, evidence of time-dependent heating was observed, an example of which can be seen in Fig.~\ref{fig:freqFluctuations}.

Here, the MD resonance frequency can be seen to vary in time at a level of around 1 kHz. Note that the corresponding shift of the $\mathrm{S}_{2}$ frequency could not be observed in this measurement since it would be smaller than the frequency resolution of this measurement. We emphasize that this observation of frequency instability serves as just an example: in general the magnitude and nature of the noise varied from shot to shot. This oscillation in frequency corresponds to a power fluctuation that exceeds the stability of the heating beam servo. We attribute this excess noise either to beam pointing of the heating beam or to internal thermal effects of the absorber itself. 

This instability should not affect the results of steady state heating measurements of a single non-hybridized mode. However, this fluctuation manifests itself in a time-dependent detuning --- and thus a time-dependent hybridization --- between the MD and $\mathrm{S}_{2}$ modes. Modeling of the full dynamics associated with this effect was not pursued in this work. However, we believe this effect explains the discrepancy between experimental results displayed in Fig.~\ref{Fig: Hybridization_results} and theoretical prediction of $\approx$ 750 K from Eq.~\ref{eq: effective_two_modes_temperature_limits}.

\bibliography{bibliography.bib}

\begin{thebibliography}{44}%
\makeatletter
\providecommand \@ifxundefined [1]{%
 \@ifx{#1\undefined}
}%
\providecommand \@ifnum [1]{%
 \ifnum #1\expandafter \@firstoftwo
 \else \expandafter \@secondoftwo
 \fi
}%
\providecommand \@ifx [1]{%
 \ifx #1\expandafter \@firstoftwo
 \else \expandafter \@secondoftwo
 \fi
}%
\providecommand \natexlab [1]{#1}%
\providecommand \enquote  [1]{``#1''}%
\providecommand \bibnamefont  [1]{#1}%
\providecommand \bibfnamefont [1]{#1}%
\providecommand \citenamefont [1]{#1}%
\providecommand \href@noop [0]{\@secondoftwo}%
\providecommand \href [0]{\begingroup \@sanitize@url \@href}%
\providecommand \@href[1]{\@@startlink{#1}\@@href}%
\providecommand \@@href[1]{\endgroup#1\@@endlink}%
\providecommand \@sanitize@url [0]{\catcode `\\12\catcode `\$12\catcode
  `\&12\catcode `\#12\catcode `\^12\catcode `\_12\catcode `\%12\relax}%
\providecommand \@@startlink[1]{}%
\providecommand \@@endlink[0]{}%
\providecommand \url  [0]{\begingroup\@sanitize@url \@url }%
\providecommand \@url [1]{\endgroup\@href {#1}{\urlprefix }}%
\providecommand \urlprefix  [0]{URL }%
\providecommand \Eprint [0]{\href }%
\providecommand \doibase [0]{http://dx.doi.org/}%
\providecommand \selectlanguage [0]{\@gobble}%
\providecommand \bibinfo  [0]{\@secondoftwo}%
\providecommand \bibfield  [0]{\@secondoftwo}%
\providecommand \translation [1]{[#1]}%
\providecommand \BibitemOpen [0]{}%
\providecommand \bibitemStop [0]{}%
\providecommand \bibitemNoStop [0]{.\EOS\space}%
\providecommand \EOS [0]{\spacefactor3000\relax}%
\providecommand \BibitemShut  [1]{\csname bibitem#1\endcsname}%
\let\auto@bib@innerbib\@empty
\bibitem [{\citenamefont {Passaro}\ \emph {et~al.}(2017)\citenamefont
  {Passaro}, \citenamefont {Cuccovillo}, \citenamefont {Vaiani}, \citenamefont
  {De~Carlo},\ and\ \citenamefont {Campanella}}]{passaro2017gyroscope}%
  \BibitemOpen
  \bibfield  {author} {\bibinfo {author} {\bibfnamefont {V.~M.}\ \bibnamefont
  {Passaro}}, \bibinfo {author} {\bibfnamefont {A.}~\bibnamefont {Cuccovillo}},
  \bibinfo {author} {\bibfnamefont {L.}~\bibnamefont {Vaiani}}, \bibinfo
  {author} {\bibfnamefont {M.}~\bibnamefont {De~Carlo}}, \ and\ \bibinfo
  {author} {\bibfnamefont {C.~E.}\ \bibnamefont {Campanella}},\ }\href@noop {}
  {\bibfield  {journal} {\bibinfo  {journal} {Sensors}\ }\textbf {\bibinfo
  {volume} {17}},\ \bibinfo {pages} {2284} (\bibinfo {year}
  {2017})}\BibitemShut {NoStop}%
\bibitem [{\citenamefont {Krause}\ \emph {et~al.}(2012)\citenamefont {Krause},
  \citenamefont {Winger}, \citenamefont {Blasius}, \citenamefont {Lin},\ and\
  \citenamefont {Painter}}]{krause2012high}%
  \BibitemOpen
  \bibfield  {author} {\bibinfo {author} {\bibfnamefont {A.~G.}\ \bibnamefont
  {Krause}}, \bibinfo {author} {\bibfnamefont {M.}~\bibnamefont {Winger}},
  \bibinfo {author} {\bibfnamefont {T.~D.}\ \bibnamefont {Blasius}}, \bibinfo
  {author} {\bibfnamefont {Q.}~\bibnamefont {Lin}}, \ and\ \bibinfo {author}
  {\bibfnamefont {O.}~\bibnamefont {Painter}},\ }\href@noop {} {\bibfield
  {journal} {\bibinfo  {journal} {Nature Photonics}\ }\textbf {\bibinfo
  {volume} {6}},\ \bibinfo {pages} {768} (\bibinfo {year} {2012})}\BibitemShut
  {NoStop}%
\bibitem [{\citenamefont {Giessibl}(2003)}]{giessibl2003advances}%
  \BibitemOpen
  \bibfield  {author} {\bibinfo {author} {\bibfnamefont {F.~J.}\ \bibnamefont
  {Giessibl}},\ }\href@noop {} {\bibfield  {journal} {\bibinfo  {journal}
  {Reviews of modern physics}\ }\textbf {\bibinfo {volume} {75}},\ \bibinfo
  {pages} {949} (\bibinfo {year} {2003})}\BibitemShut {NoStop}%
\bibitem [{\citenamefont {Poggio}\ and\ \citenamefont
  {Degen}(2010)}]{poggio2010force}%
  \BibitemOpen
  \bibfield  {author} {\bibinfo {author} {\bibfnamefont {M.}~\bibnamefont
  {Poggio}}\ and\ \bibinfo {author} {\bibfnamefont {C.~L.}\ \bibnamefont
  {Degen}},\ }\href@noop {} {\bibfield  {journal} {\bibinfo  {journal}
  {Nanotechnology}\ }\textbf {\bibinfo {volume} {21}},\ \bibinfo {pages}
  {342001} (\bibinfo {year} {2010})}\BibitemShut {NoStop}%
\bibitem [{\citenamefont {Degen}\ \emph {et~al.}(2009)\citenamefont {Degen},
  \citenamefont {Poggio}, \citenamefont {Mamin}, \citenamefont {Rettner},\ and\
  \citenamefont {Rugar}}]{degen2009nanoscale}%
  \BibitemOpen
  \bibfield  {author} {\bibinfo {author} {\bibfnamefont {C.}~\bibnamefont
  {Degen}}, \bibinfo {author} {\bibfnamefont {M.}~\bibnamefont {Poggio}},
  \bibinfo {author} {\bibfnamefont {H.}~\bibnamefont {Mamin}}, \bibinfo
  {author} {\bibfnamefont {C.}~\bibnamefont {Rettner}}, \ and\ \bibinfo
  {author} {\bibfnamefont {D.}~\bibnamefont {Rugar}},\ }\href@noop {}
  {\bibfield  {journal} {\bibinfo  {journal} {Proceedings of the National
  Academy of Sciences}\ }\textbf {\bibinfo {volume} {106}},\ \bibinfo {pages}
  {1313} (\bibinfo {year} {2009})}\BibitemShut {NoStop}%
\bibitem [{\citenamefont {Arlett}\ \emph {et~al.}(2011)\citenamefont {Arlett},
  \citenamefont {Myers},\ and\ \citenamefont {Roukes}}]{arlett2011comparative}%
  \BibitemOpen
  \bibfield  {author} {\bibinfo {author} {\bibfnamefont {J.}~\bibnamefont
  {Arlett}}, \bibinfo {author} {\bibfnamefont {E.}~\bibnamefont {Myers}}, \
  and\ \bibinfo {author} {\bibfnamefont {M.}~\bibnamefont {Roukes}},\
  }\href@noop {} {\bibfield  {journal} {\bibinfo  {journal} {Nature
  nanotechnology}\ }\textbf {\bibinfo {volume} {6}},\ \bibinfo {pages} {203}
  (\bibinfo {year} {2011})}\BibitemShut {NoStop}%
\bibitem [{\citenamefont {Bleszynski-Jayich}\ \emph {et~al.}(2009)\citenamefont
  {Bleszynski-Jayich}, \citenamefont {Shanks}, \citenamefont {Peaudecerf},
  \citenamefont {Ginossar}, \citenamefont {Von~Oppen}, \citenamefont
  {Glazman},\ and\ \citenamefont {Harris}}]{bleszynski2009persistent}%
  \BibitemOpen
  \bibfield  {author} {\bibinfo {author} {\bibfnamefont {A.}~\bibnamefont
  {Bleszynski-Jayich}}, \bibinfo {author} {\bibfnamefont {W.}~\bibnamefont
  {Shanks}}, \bibinfo {author} {\bibfnamefont {B.}~\bibnamefont {Peaudecerf}},
  \bibinfo {author} {\bibfnamefont {E.}~\bibnamefont {Ginossar}}, \bibinfo
  {author} {\bibfnamefont {F.}~\bibnamefont {Von~Oppen}}, \bibinfo {author}
  {\bibfnamefont {L.}~\bibnamefont {Glazman}}, \ and\ \bibinfo {author}
  {\bibfnamefont {J.}~\bibnamefont {Harris}},\ }\href@noop {} {\bibfield
  {journal} {\bibinfo  {journal} {Science}\ }\textbf {\bibinfo {volume}
  {326}},\ \bibinfo {pages} {272} (\bibinfo {year} {2009})}\BibitemShut
  {NoStop}%
\bibitem [{\citenamefont {Andrews}\ \emph {et~al.}(2014)\citenamefont
  {Andrews}, \citenamefont {Peterson}, \citenamefont {Purdy}, \citenamefont
  {Cicak}, \citenamefont {Simmonds}, \citenamefont {Regal},\ and\ \citenamefont
  {Lehnert}}]{andrews2014bidirectional}%
  \BibitemOpen
  \bibfield  {author} {\bibinfo {author} {\bibfnamefont {R.~W.}\ \bibnamefont
  {Andrews}}, \bibinfo {author} {\bibfnamefont {R.~W.}\ \bibnamefont
  {Peterson}}, \bibinfo {author} {\bibfnamefont {T.~P.}\ \bibnamefont {Purdy}},
  \bibinfo {author} {\bibfnamefont {K.}~\bibnamefont {Cicak}}, \bibinfo
  {author} {\bibfnamefont {R.~W.}\ \bibnamefont {Simmonds}}, \bibinfo {author}
  {\bibfnamefont {C.~A.}\ \bibnamefont {Regal}}, \ and\ \bibinfo {author}
  {\bibfnamefont {K.~W.}\ \bibnamefont {Lehnert}},\ }\href@noop {} {\bibfield
  {journal} {\bibinfo  {journal} {Nature Physics}\ }\textbf {\bibinfo {volume}
  {10}},\ \bibinfo {pages} {321} (\bibinfo {year} {2014})}\BibitemShut
  {NoStop}%
\bibitem [{\citenamefont {Wallucks}\ \emph {et~al.}(2020)\citenamefont
  {Wallucks}, \citenamefont {Marinkovi{\'c}}, \citenamefont {Hensen},
  \citenamefont {Stockill},\ and\ \citenamefont
  {Gr{\"o}blacher}}]{wallucks2020quantum}%
  \BibitemOpen
  \bibfield  {author} {\bibinfo {author} {\bibfnamefont {A.}~\bibnamefont
  {Wallucks}}, \bibinfo {author} {\bibfnamefont {I.}~\bibnamefont
  {Marinkovi{\'c}}}, \bibinfo {author} {\bibfnamefont {B.}~\bibnamefont
  {Hensen}}, \bibinfo {author} {\bibfnamefont {R.}~\bibnamefont {Stockill}}, \
  and\ \bibinfo {author} {\bibfnamefont {S.}~\bibnamefont {Gr{\"o}blacher}},\
  }\href@noop {} {\bibfield  {journal} {\bibinfo  {journal} {Nature Physics}\
  }\textbf {\bibinfo {volume} {16}},\ \bibinfo {pages} {772} (\bibinfo {year}
  {2020})}\BibitemShut {NoStop}%
\bibitem [{\citenamefont {Barzanjeh}\ \emph {et~al.}(2022)\citenamefont
  {Barzanjeh}, \citenamefont {Xuereb}, \citenamefont {Gr{\"o}blacher},
  \citenamefont {Paternostro}, \citenamefont {Regal},\ and\ \citenamefont
  {Weig}}]{barzanjeh2022optomechanics}%
  \BibitemOpen
  \bibfield  {author} {\bibinfo {author} {\bibfnamefont {S.}~\bibnamefont
  {Barzanjeh}}, \bibinfo {author} {\bibfnamefont {A.}~\bibnamefont {Xuereb}},
  \bibinfo {author} {\bibfnamefont {S.}~\bibnamefont {Gr{\"o}blacher}},
  \bibinfo {author} {\bibfnamefont {M.}~\bibnamefont {Paternostro}}, \bibinfo
  {author} {\bibfnamefont {C.~A.}\ \bibnamefont {Regal}}, \ and\ \bibinfo
  {author} {\bibfnamefont {E.~M.}\ \bibnamefont {Weig}},\ }\href@noop {}
  {\bibfield  {journal} {\bibinfo  {journal} {Nature Physics}\ }\textbf
  {\bibinfo {volume} {18}},\ \bibinfo {pages} {15} (\bibinfo {year}
  {2022})}\BibitemShut {NoStop}%
\bibitem [{\citenamefont {Schm{\"o}le}\ \emph {et~al.}(2016)\citenamefont
  {Schm{\"o}le}, \citenamefont {Dragosits}, \citenamefont {Hepach},\ and\
  \citenamefont {Aspelmeyer}}]{schmole2016micromechanical}%
  \BibitemOpen
  \bibfield  {author} {\bibinfo {author} {\bibfnamefont {J.}~\bibnamefont
  {Schm{\"o}le}}, \bibinfo {author} {\bibfnamefont {M.}~\bibnamefont
  {Dragosits}}, \bibinfo {author} {\bibfnamefont {H.}~\bibnamefont {Hepach}}, \
  and\ \bibinfo {author} {\bibfnamefont {M.}~\bibnamefont {Aspelmeyer}},\
  }\href@noop {} {\bibfield  {journal} {\bibinfo  {journal} {Classical and
  Quantum Gravity}\ }\textbf {\bibinfo {volume} {33}},\ \bibinfo {pages}
  {125031} (\bibinfo {year} {2016})}\BibitemShut {NoStop}%
\bibitem [{\citenamefont {Liu}\ \emph {et~al.}(2021)\citenamefont {Liu},
  \citenamefont {Mummery}, \citenamefont {Zhou},\ and\ \citenamefont
  {Sillanp{\"a}{\"a}}}]{liu2021gravitational}%
  \BibitemOpen
  \bibfield  {author} {\bibinfo {author} {\bibfnamefont {Y.}~\bibnamefont
  {Liu}}, \bibinfo {author} {\bibfnamefont {J.}~\bibnamefont {Mummery}},
  \bibinfo {author} {\bibfnamefont {J.}~\bibnamefont {Zhou}}, \ and\ \bibinfo
  {author} {\bibfnamefont {M.~A.}\ \bibnamefont {Sillanp{\"a}{\"a}}},\
  }\href@noop {} {\bibfield  {journal} {\bibinfo  {journal} {Physical Review
  Applied}\ }\textbf {\bibinfo {volume} {15}},\ \bibinfo {pages} {034004}
  (\bibinfo {year} {2021})}\BibitemShut {NoStop}%
\bibitem [{\citenamefont {Carney}\ \emph {et~al.}(2021)\citenamefont {Carney},
  \citenamefont {Krnjaic}, \citenamefont {Moore}, \citenamefont {Regal},
  \citenamefont {Afek}, \citenamefont {Bhave}, \citenamefont {Brubaker},
  \citenamefont {Corbitt}, \citenamefont {Cripe}, \citenamefont {Crisosto}
  \emph {et~al.}}]{carney2021mechanical}%
  \BibitemOpen
  \bibfield  {author} {\bibinfo {author} {\bibfnamefont {D.}~\bibnamefont
  {Carney}}, \bibinfo {author} {\bibfnamefont {G.}~\bibnamefont {Krnjaic}},
  \bibinfo {author} {\bibfnamefont {D.~C.}\ \bibnamefont {Moore}}, \bibinfo
  {author} {\bibfnamefont {C.~A.}\ \bibnamefont {Regal}}, \bibinfo {author}
  {\bibfnamefont {G.}~\bibnamefont {Afek}}, \bibinfo {author} {\bibfnamefont
  {S.}~\bibnamefont {Bhave}}, \bibinfo {author} {\bibfnamefont
  {B.}~\bibnamefont {Brubaker}}, \bibinfo {author} {\bibfnamefont
  {T.}~\bibnamefont {Corbitt}}, \bibinfo {author} {\bibfnamefont
  {J.}~\bibnamefont {Cripe}}, \bibinfo {author} {\bibfnamefont
  {N.}~\bibnamefont {Crisosto}},  \emph {et~al.},\ }\href@noop {} {\bibfield
  {journal} {\bibinfo  {journal} {Quantum Science and Technology}\ }\textbf
  {\bibinfo {volume} {6}},\ \bibinfo {pages} {024002} (\bibinfo {year}
  {2021})}\BibitemShut {NoStop}%
\bibitem [{\citenamefont {Piller}\ \emph {et~al.}(2022)\citenamefont {Piller},
  \citenamefont {Hiesberger}, \citenamefont {Wistrela}, \citenamefont
  {Martini}, \citenamefont {Luhmann},\ and\ \citenamefont
  {Schmid}}]{piller2022thermal}%
  \BibitemOpen
  \bibfield  {author} {\bibinfo {author} {\bibfnamefont {M.}~\bibnamefont
  {Piller}}, \bibinfo {author} {\bibfnamefont {J.}~\bibnamefont {Hiesberger}},
  \bibinfo {author} {\bibfnamefont {E.}~\bibnamefont {Wistrela}}, \bibinfo
  {author} {\bibfnamefont {P.}~\bibnamefont {Martini}}, \bibinfo {author}
  {\bibfnamefont {N.}~\bibnamefont {Luhmann}}, \ and\ \bibinfo {author}
  {\bibfnamefont {S.}~\bibnamefont {Schmid}},\ }\href@noop {} {\bibfield
  {journal} {\bibinfo  {journal} {IEEE Sensors Journal}\ } (\bibinfo {year}
  {2022})}\BibitemShut {NoStop}%
\bibitem [{\citenamefont {Harry}\ \emph {et~al.}(2010)\citenamefont {Harry},
  \citenamefont {for~the LIGO Scientific~Collaboration} \emph
  {et~al.}}]{harry2010advanced}%
  \BibitemOpen
  \bibfield  {author} {\bibinfo {author} {\bibfnamefont {G.~M.}\ \bibnamefont
  {Harry}}, \bibinfo {author} {\bibnamefont {for~the LIGO
  Scientific~Collaboration}},  \emph {et~al.},\ }\href@noop {} {\bibfield
  {journal} {\bibinfo  {journal} {Classical and Quantum Gravity}\ }\textbf
  {\bibinfo {volume} {27}},\ \bibinfo {pages} {084006} (\bibinfo {year}
  {2010})}\BibitemShut {NoStop}%
\bibitem [{\citenamefont {Riedinger}\ \emph {et~al.}(2018)\citenamefont
  {Riedinger}, \citenamefont {Wallucks}, \citenamefont {Marinkovi{\'c}},
  \citenamefont {L{\"o}schnauer}, \citenamefont {Aspelmeyer}, \citenamefont
  {Hong},\ and\ \citenamefont {Gr{\"o}blacher}}]{riedinger2018remote}%
  \BibitemOpen
  \bibfield  {author} {\bibinfo {author} {\bibfnamefont {R.}~\bibnamefont
  {Riedinger}}, \bibinfo {author} {\bibfnamefont {A.}~\bibnamefont {Wallucks}},
  \bibinfo {author} {\bibfnamefont {I.}~\bibnamefont {Marinkovi{\'c}}},
  \bibinfo {author} {\bibfnamefont {C.}~\bibnamefont {L{\"o}schnauer}},
  \bibinfo {author} {\bibfnamefont {M.}~\bibnamefont {Aspelmeyer}}, \bibinfo
  {author} {\bibfnamefont {S.}~\bibnamefont {Hong}}, \ and\ \bibinfo {author}
  {\bibfnamefont {S.}~\bibnamefont {Gr{\"o}blacher}},\ }\href@noop {}
  {\bibfield  {journal} {\bibinfo  {journal} {Nature}\ }\textbf {\bibinfo
  {volume} {556}},\ \bibinfo {pages} {473} (\bibinfo {year}
  {2018})}\BibitemShut {NoStop}%
\bibitem [{\citenamefont {Mirhosseini}\ \emph {et~al.}(2020)\citenamefont
  {Mirhosseini}, \citenamefont {Sipahigil}, \citenamefont {Kalaee},\ and\
  \citenamefont {Painter}}]{mirhosseini2020superconducting}%
  \BibitemOpen
  \bibfield  {author} {\bibinfo {author} {\bibfnamefont {M.}~\bibnamefont
  {Mirhosseini}}, \bibinfo {author} {\bibfnamefont {A.}~\bibnamefont
  {Sipahigil}}, \bibinfo {author} {\bibfnamefont {M.}~\bibnamefont {Kalaee}}, \
  and\ \bibinfo {author} {\bibfnamefont {O.}~\bibnamefont {Painter}},\
  }\href@noop {} {\bibfield  {journal} {\bibinfo  {journal} {Nature}\ }\textbf
  {\bibinfo {volume} {588}},\ \bibinfo {pages} {599} (\bibinfo {year}
  {2020})}\BibitemShut {NoStop}%
\bibitem [{\citenamefont {Qiu}\ \emph {et~al.}(2020)\citenamefont {Qiu},
  \citenamefont {Shomroni}, \citenamefont {Seidler},\ and\ \citenamefont
  {Kippenberg}}]{qiu2020laser}%
  \BibitemOpen
  \bibfield  {author} {\bibinfo {author} {\bibfnamefont {L.}~\bibnamefont
  {Qiu}}, \bibinfo {author} {\bibfnamefont {I.}~\bibnamefont {Shomroni}},
  \bibinfo {author} {\bibfnamefont {P.}~\bibnamefont {Seidler}}, \ and\
  \bibinfo {author} {\bibfnamefont {T.~J.}\ \bibnamefont {Kippenberg}},\
  }\href@noop {} {\bibfield  {journal} {\bibinfo  {journal} {Physical Review
  Letters}\ }\textbf {\bibinfo {volume} {124}},\ \bibinfo {pages} {173601}
  (\bibinfo {year} {2020})}\BibitemShut {NoStop}%
\bibitem [{\citenamefont {Peterson}\ \emph {et~al.}(2016)\citenamefont
  {Peterson}, \citenamefont {Purdy}, \citenamefont {Kampel}, \citenamefont
  {Andrews}, \citenamefont {Yu}, \citenamefont {Lehnert},\ and\ \citenamefont
  {Regal}}]{peterson2016laser}%
  \BibitemOpen
  \bibfield  {author} {\bibinfo {author} {\bibfnamefont {R.}~\bibnamefont
  {Peterson}}, \bibinfo {author} {\bibfnamefont {T.}~\bibnamefont {Purdy}},
  \bibinfo {author} {\bibfnamefont {N.}~\bibnamefont {Kampel}}, \bibinfo
  {author} {\bibfnamefont {R.}~\bibnamefont {Andrews}}, \bibinfo {author}
  {\bibfnamefont {P.-L.}\ \bibnamefont {Yu}}, \bibinfo {author} {\bibfnamefont
  {K.}~\bibnamefont {Lehnert}}, \ and\ \bibinfo {author} {\bibfnamefont
  {C.}~\bibnamefont {Regal}},\ }\href@noop {} {\bibfield  {journal} {\bibinfo
  {journal} {Physical review letters}\ }\textbf {\bibinfo {volume} {116}},\
  \bibinfo {pages} {063601} (\bibinfo {year} {2016})}\BibitemShut {NoStop}%
\bibitem [{\citenamefont {Page}\ \emph {et~al.}(2021)\citenamefont {Page},
  \citenamefont {Goryachev}, \citenamefont {Miao}, \citenamefont {Chen},
  \citenamefont {Ma}, \citenamefont {Mason}, \citenamefont {Rossi},
  \citenamefont {Blair}, \citenamefont {Ju}, \citenamefont {Blair} \emph
  {et~al.}}]{page2021gravitational}%
  \BibitemOpen
  \bibfield  {author} {\bibinfo {author} {\bibfnamefont {M.~A.}\ \bibnamefont
  {Page}}, \bibinfo {author} {\bibfnamefont {M.}~\bibnamefont {Goryachev}},
  \bibinfo {author} {\bibfnamefont {H.}~\bibnamefont {Miao}}, \bibinfo {author}
  {\bibfnamefont {Y.}~\bibnamefont {Chen}}, \bibinfo {author} {\bibfnamefont
  {Y.}~\bibnamefont {Ma}}, \bibinfo {author} {\bibfnamefont {D.}~\bibnamefont
  {Mason}}, \bibinfo {author} {\bibfnamefont {M.}~\bibnamefont {Rossi}},
  \bibinfo {author} {\bibfnamefont {C.~D.}\ \bibnamefont {Blair}}, \bibinfo
  {author} {\bibfnamefont {L.}~\bibnamefont {Ju}}, \bibinfo {author}
  {\bibfnamefont {D.~G.}\ \bibnamefont {Blair}},  \emph {et~al.},\ }\href@noop
  {} {\bibfield  {journal} {\bibinfo  {journal} {Communications Physics}\
  }\textbf {\bibinfo {volume} {4}},\ \bibinfo {pages} {27} (\bibinfo {year}
  {2021})}\BibitemShut {NoStop}%
\bibitem [{\citenamefont {Lax}(1960)}]{lax1960fluctuations}%
  \BibitemOpen
  \bibfield  {author} {\bibinfo {author} {\bibfnamefont {M.}~\bibnamefont
  {Lax}},\ }\href@noop {} {\bibfield  {journal} {\bibinfo  {journal} {Reviews
  of modern physics}\ }\textbf {\bibinfo {volume} {32}},\ \bibinfo {pages} {25}
  (\bibinfo {year} {1960})}\BibitemShut {NoStop}%
\bibitem [{\citenamefont {Komori}\ \emph {et~al.}(2018)\citenamefont {Komori},
  \citenamefont {Enomoto}, \citenamefont {Takeda}, \citenamefont {Michimura},
  \citenamefont {Somiya}, \citenamefont {Ando},\ and\ \citenamefont
  {Ballmer}}]{komori2018direct}%
  \BibitemOpen
  \bibfield  {author} {\bibinfo {author} {\bibfnamefont {K.}~\bibnamefont
  {Komori}}, \bibinfo {author} {\bibfnamefont {Y.}~\bibnamefont {Enomoto}},
  \bibinfo {author} {\bibfnamefont {H.}~\bibnamefont {Takeda}}, \bibinfo
  {author} {\bibfnamefont {Y.}~\bibnamefont {Michimura}}, \bibinfo {author}
  {\bibfnamefont {K.}~\bibnamefont {Somiya}}, \bibinfo {author} {\bibfnamefont
  {M.}~\bibnamefont {Ando}}, \ and\ \bibinfo {author} {\bibfnamefont {S.~W.}\
  \bibnamefont {Ballmer}},\ }\href@noop {} {\bibfield  {journal} {\bibinfo
  {journal} {Physical Review D}\ }\textbf {\bibinfo {volume} {97}},\ \bibinfo
  {pages} {102001} (\bibinfo {year} {2018})}\BibitemShut {NoStop}%
\bibitem [{\citenamefont {Geitner}\ \emph {et~al.}(2017)\citenamefont
  {Geitner}, \citenamefont {Sandoval}, \citenamefont {Bertin},\ and\
  \citenamefont {Bellon}}]{geitner2017low}%
  \BibitemOpen
  \bibfield  {author} {\bibinfo {author} {\bibfnamefont {M.}~\bibnamefont
  {Geitner}}, \bibinfo {author} {\bibfnamefont {F.~A.}\ \bibnamefont
  {Sandoval}}, \bibinfo {author} {\bibfnamefont {E.}~\bibnamefont {Bertin}}, \
  and\ \bibinfo {author} {\bibfnamefont {L.}~\bibnamefont {Bellon}},\
  }\href@noop {} {\bibfield  {journal} {\bibinfo  {journal} {Physical Review
  E}\ }\textbf {\bibinfo {volume} {95}},\ \bibinfo {pages} {032138} (\bibinfo
  {year} {2017})}\BibitemShut {NoStop}%
\bibitem [{\citenamefont {Singh}\ and\ \citenamefont
  {Purdy}(2020)}]{singh2020detecting}%
  \BibitemOpen
  \bibfield  {author} {\bibinfo {author} {\bibfnamefont {R.}~\bibnamefont
  {Singh}}\ and\ \bibinfo {author} {\bibfnamefont {T.~P.}\ \bibnamefont
  {Purdy}},\ }\href@noop {} {\bibfield  {journal} {\bibinfo  {journal}
  {Physical Review Letters}\ }\textbf {\bibinfo {volume} {125}},\ \bibinfo
  {pages} {120603} (\bibinfo {year} {2020})}\BibitemShut {NoStop}%
\bibitem [{\citenamefont {Ghadimi}\ \emph {et~al.}(2018)\citenamefont
  {Ghadimi}, \citenamefont {Fedorov}, \citenamefont {Engelsen}, \citenamefont
  {Bereyhi}, \citenamefont {Schilling}, \citenamefont {Wilson},\ and\
  \citenamefont {Kippenberg}}]{ghadimi2018elastic}%
  \BibitemOpen
  \bibfield  {author} {\bibinfo {author} {\bibfnamefont {A.~H.}\ \bibnamefont
  {Ghadimi}}, \bibinfo {author} {\bibfnamefont {S.~A.}\ \bibnamefont
  {Fedorov}}, \bibinfo {author} {\bibfnamefont {N.~J.}\ \bibnamefont
  {Engelsen}}, \bibinfo {author} {\bibfnamefont {M.~J.}\ \bibnamefont
  {Bereyhi}}, \bibinfo {author} {\bibfnamefont {R.}~\bibnamefont {Schilling}},
  \bibinfo {author} {\bibfnamefont {D.~J.}\ \bibnamefont {Wilson}}, \ and\
  \bibinfo {author} {\bibfnamefont {T.~J.}\ \bibnamefont {Kippenberg}},\
  }\href@noop {} {\bibfield  {journal} {\bibinfo  {journal} {Science}\ }\textbf
  {\bibinfo {volume} {360}},\ \bibinfo {pages} {764} (\bibinfo {year}
  {2018})}\BibitemShut {NoStop}%
\bibitem [{\citenamefont {Fedorov}\ \emph {et~al.}(2020)\citenamefont
  {Fedorov}, \citenamefont {Beccari}, \citenamefont {Engelsen},\ and\
  \citenamefont {Kippenberg}}]{fedorov2020fractal}%
  \BibitemOpen
  \bibfield  {author} {\bibinfo {author} {\bibfnamefont {S.~A.}\ \bibnamefont
  {Fedorov}}, \bibinfo {author} {\bibfnamefont {A.}~\bibnamefont {Beccari}},
  \bibinfo {author} {\bibfnamefont {N.~J.}\ \bibnamefont {Engelsen}}, \ and\
  \bibinfo {author} {\bibfnamefont {T.~J.}\ \bibnamefont {Kippenberg}},\
  }\href@noop {} {\bibfield  {journal} {\bibinfo  {journal} {Physical Review
  Letters}\ }\textbf {\bibinfo {volume} {124}},\ \bibinfo {pages} {025502}
  (\bibinfo {year} {2020})}\BibitemShut {NoStop}%
\bibitem [{\citenamefont {H{\o}j}\ \emph {et~al.}(2021)\citenamefont {H{\o}j},
  \citenamefont {Wang}, \citenamefont {Gao}, \citenamefont {Hoff},
  \citenamefont {Sigmund},\ and\ \citenamefont {Andersen}}]{hoj2021ultra}%
  \BibitemOpen
  \bibfield  {author} {\bibinfo {author} {\bibfnamefont {D.}~\bibnamefont
  {H{\o}j}}, \bibinfo {author} {\bibfnamefont {F.}~\bibnamefont {Wang}},
  \bibinfo {author} {\bibfnamefont {W.}~\bibnamefont {Gao}}, \bibinfo {author}
  {\bibfnamefont {U.~B.}\ \bibnamefont {Hoff}}, \bibinfo {author}
  {\bibfnamefont {O.}~\bibnamefont {Sigmund}}, \ and\ \bibinfo {author}
  {\bibfnamefont {U.~L.}\ \bibnamefont {Andersen}},\ }\href@noop {} {\bibfield
  {journal} {\bibinfo  {journal} {Nature communications}\ }\textbf {\bibinfo
  {volume} {12}},\ \bibinfo {pages} {5766} (\bibinfo {year}
  {2021})}\BibitemShut {NoStop}%
\bibitem [{\citenamefont {Bereyhi}\ \emph {et~al.}(2022)\citenamefont
  {Bereyhi}, \citenamefont {Arabmoheghi}, \citenamefont {Beccari},
  \citenamefont {Fedorov}, \citenamefont {Huang}, \citenamefont {Kippenberg},\
  and\ \citenamefont {Engelsen}}]{bereyhi2022perimeter}%
  \BibitemOpen
  \bibfield  {author} {\bibinfo {author} {\bibfnamefont {M.~J.}\ \bibnamefont
  {Bereyhi}}, \bibinfo {author} {\bibfnamefont {A.}~\bibnamefont
  {Arabmoheghi}}, \bibinfo {author} {\bibfnamefont {A.}~\bibnamefont
  {Beccari}}, \bibinfo {author} {\bibfnamefont {S.~A.}\ \bibnamefont
  {Fedorov}}, \bibinfo {author} {\bibfnamefont {G.}~\bibnamefont {Huang}},
  \bibinfo {author} {\bibfnamefont {T.~J.}\ \bibnamefont {Kippenberg}}, \ and\
  \bibinfo {author} {\bibfnamefont {N.~J.}\ \bibnamefont {Engelsen}},\
  }\href@noop {} {\bibfield  {journal} {\bibinfo  {journal} {Physical Review
  X}\ }\textbf {\bibinfo {volume} {12}},\ \bibinfo {pages} {021036} (\bibinfo
  {year} {2022})}\BibitemShut {NoStop}%
\bibitem [{\citenamefont {Pluchar}\ \emph {et~al.}(2022)\citenamefont
  {Pluchar}, \citenamefont {Agrawal}, \citenamefont {Condos}, \citenamefont
  {Pratt}, \citenamefont {Schlamminger},\ and\ \citenamefont
  {Wilson}}]{pluchar2022high}%
  \BibitemOpen
  \bibfield  {author} {\bibinfo {author} {\bibfnamefont {C.~M.}\ \bibnamefont
  {Pluchar}}, \bibinfo {author} {\bibfnamefont {A.~R.}\ \bibnamefont
  {Agrawal}}, \bibinfo {author} {\bibfnamefont {C.~A.}\ \bibnamefont {Condos}},
  \bibinfo {author} {\bibfnamefont {J.}~\bibnamefont {Pratt}}, \bibinfo
  {author} {\bibfnamefont {S.}~\bibnamefont {Schlamminger}}, \ and\ \bibinfo
  {author} {\bibfnamefont {D.}~\bibnamefont {Wilson}},\ }in\ \href@noop {}
  {\emph {\bibinfo {booktitle} {Optical Trapping and Optical Micromanipulation
  XIX}}}\ (\bibinfo {organization} {SPIE},\ \bibinfo {year} {2022})\ p.\
  \bibinfo {pages} {PC121980U}\BibitemShut {NoStop}%
\bibitem [{\citenamefont {Eichenfield}\ \emph {et~al.}(2009)\citenamefont
  {Eichenfield}, \citenamefont {Chan}, \citenamefont {Camacho}, \citenamefont
  {Vahala},\ and\ \citenamefont {Painter}}]{eichenfield2009optomechanical}%
  \BibitemOpen
  \bibfield  {author} {\bibinfo {author} {\bibfnamefont {M.}~\bibnamefont
  {Eichenfield}}, \bibinfo {author} {\bibfnamefont {J.}~\bibnamefont {Chan}},
  \bibinfo {author} {\bibfnamefont {R.~M.}\ \bibnamefont {Camacho}}, \bibinfo
  {author} {\bibfnamefont {K.~J.}\ \bibnamefont {Vahala}}, \ and\ \bibinfo
  {author} {\bibfnamefont {O.}~\bibnamefont {Painter}},\ }\href@noop {}
  {\bibfield  {journal} {\bibinfo  {journal} {nature}\ }\textbf {\bibinfo
  {volume} {462}},\ \bibinfo {pages} {78} (\bibinfo {year} {2009})}\BibitemShut
  {NoStop}%
\bibitem [{\citenamefont {Clerk}\ \emph {et~al.}(2010)\citenamefont {Clerk},
  \citenamefont {Devoret}, \citenamefont {Girvin}, \citenamefont {Marquardt},\
  and\ \citenamefont {Schoelkopf}}]{clerk2010introduction}%
  \BibitemOpen
  \bibfield  {author} {\bibinfo {author} {\bibfnamefont {A.~A.}\ \bibnamefont
  {Clerk}}, \bibinfo {author} {\bibfnamefont {M.~H.}\ \bibnamefont {Devoret}},
  \bibinfo {author} {\bibfnamefont {S.~M.}\ \bibnamefont {Girvin}}, \bibinfo
  {author} {\bibfnamefont {F.}~\bibnamefont {Marquardt}}, \ and\ \bibinfo
  {author} {\bibfnamefont {R.~J.}\ \bibnamefont {Schoelkopf}},\ }\href@noop {}
  {\bibfield  {journal} {\bibinfo  {journal} {Reviews of Modern Physics}\
  }\textbf {\bibinfo {volume} {82}},\ \bibinfo {pages} {1155} (\bibinfo {year}
  {2010})}\BibitemShut {NoStop}%
\bibitem [{\citenamefont {Kushwaha}\ \emph {et~al.}(1993)\citenamefont
  {Kushwaha}, \citenamefont {Halevi}, \citenamefont {Dobrzynski},\ and\
  \citenamefont {Djafari-Rouhani}}]{kushwaha1993acoustic}%
  \BibitemOpen
  \bibfield  {author} {\bibinfo {author} {\bibfnamefont {M.~S.}\ \bibnamefont
  {Kushwaha}}, \bibinfo {author} {\bibfnamefont {P.}~\bibnamefont {Halevi}},
  \bibinfo {author} {\bibfnamefont {L.}~\bibnamefont {Dobrzynski}}, \ and\
  \bibinfo {author} {\bibfnamefont {B.}~\bibnamefont {Djafari-Rouhani}},\
  }\href@noop {} {\bibfield  {journal} {\bibinfo  {journal} {Physical review
  letters}\ }\textbf {\bibinfo {volume} {71}},\ \bibinfo {pages} {2022}
  (\bibinfo {year} {1993})}\BibitemShut {NoStop}%
\bibitem [{\citenamefont {Tsaturyan}\ \emph {et~al.}(2017)\citenamefont
  {Tsaturyan}, \citenamefont {Barg}, \citenamefont {Polzik},\ and\
  \citenamefont {Schliesser}}]{tsaturyan2017ultracoherent}%
  \BibitemOpen
  \bibfield  {author} {\bibinfo {author} {\bibfnamefont {Y.}~\bibnamefont
  {Tsaturyan}}, \bibinfo {author} {\bibfnamefont {A.}~\bibnamefont {Barg}},
  \bibinfo {author} {\bibfnamefont {E.~S.}\ \bibnamefont {Polzik}}, \ and\
  \bibinfo {author} {\bibfnamefont {A.}~\bibnamefont {Schliesser}},\
  }\href@noop {} {\bibfield  {journal} {\bibinfo  {journal} {Nature
  nanotechnology}\ }\textbf {\bibinfo {volume} {12}},\ \bibinfo {pages} {776}
  (\bibinfo {year} {2017})}\BibitemShut {NoStop}%
\bibitem [{\citenamefont {Zhang}\ \emph {et~al.}(2020)\citenamefont {Zhang},
  \citenamefont {Giroux}, \citenamefont {Nour},\ and\ \citenamefont
  {St-Gelais}}]{zhang2020radiative}%
  \BibitemOpen
  \bibfield  {author} {\bibinfo {author} {\bibfnamefont {C.}~\bibnamefont
  {Zhang}}, \bibinfo {author} {\bibfnamefont {M.}~\bibnamefont {Giroux}},
  \bibinfo {author} {\bibfnamefont {T.~A.}\ \bibnamefont {Nour}}, \ and\
  \bibinfo {author} {\bibfnamefont {R.}~\bibnamefont {St-Gelais}},\ }\href@noop
  {} {\bibfield  {journal} {\bibinfo  {journal} {Physical Review Applied}\
  }\textbf {\bibinfo {volume} {14}},\ \bibinfo {pages} {024072} (\bibinfo
  {year} {2020})}\BibitemShut {NoStop}%
\bibitem [{\citenamefont {Leivo}\ and\ \citenamefont
  {Pekola}(1998)}]{leivo1998thermal}%
  \BibitemOpen
  \bibfield  {author} {\bibinfo {author} {\bibfnamefont {M.}~\bibnamefont
  {Leivo}}\ and\ \bibinfo {author} {\bibfnamefont {J.}~\bibnamefont {Pekola}},\
  }\href@noop {} {\bibfield  {journal} {\bibinfo  {journal} {Applied Physics
  Letters}\ }\textbf {\bibinfo {volume} {72}},\ \bibinfo {pages} {1305}
  (\bibinfo {year} {1998})}\BibitemShut {NoStop}%
\bibitem [{\citenamefont {Vicarelli}\ \emph {et~al.}(2022)\citenamefont
  {Vicarelli}, \citenamefont {Tredicucci},\ and\ \citenamefont
  {Pitanti}}]{vicarelli2022micromechanical}%
  \BibitemOpen
  \bibfield  {author} {\bibinfo {author} {\bibfnamefont {L.}~\bibnamefont
  {Vicarelli}}, \bibinfo {author} {\bibfnamefont {A.}~\bibnamefont
  {Tredicucci}}, \ and\ \bibinfo {author} {\bibfnamefont {A.}~\bibnamefont
  {Pitanti}},\ }\href@noop {} {\bibfield  {journal} {\bibinfo  {journal} {ACS
  photonics}\ }\textbf {\bibinfo {volume} {9}},\ \bibinfo {pages} {360}
  (\bibinfo {year} {2022})}\BibitemShut {NoStop}%
\bibitem [{\citenamefont {Piller}\ \emph {et~al.}(2020)\citenamefont {Piller},
  \citenamefont {Sadeghi}, \citenamefont {West}, \citenamefont {Luhmann},
  \citenamefont {Martini}, \citenamefont {Hansen},\ and\ \citenamefont
  {Schmid}}]{piller2020thermal}%
  \BibitemOpen
  \bibfield  {author} {\bibinfo {author} {\bibfnamefont {M.}~\bibnamefont
  {Piller}}, \bibinfo {author} {\bibfnamefont {P.}~\bibnamefont {Sadeghi}},
  \bibinfo {author} {\bibfnamefont {R.~G.}\ \bibnamefont {West}}, \bibinfo
  {author} {\bibfnamefont {N.}~\bibnamefont {Luhmann}}, \bibinfo {author}
  {\bibfnamefont {P.}~\bibnamefont {Martini}}, \bibinfo {author} {\bibfnamefont
  {O.}~\bibnamefont {Hansen}}, \ and\ \bibinfo {author} {\bibfnamefont
  {S.}~\bibnamefont {Schmid}},\ }\href@noop {} {\bibfield  {journal} {\bibinfo
  {journal} {Applied Physics Letters}\ }\textbf {\bibinfo {volume} {117}},\
  \bibinfo {pages} {034101} (\bibinfo {year} {2020})}\BibitemShut {NoStop}%
\bibitem [{\citenamefont {Reetz}\ \emph {et~al.}(2019)\citenamefont {Reetz},
  \citenamefont {Fischer}, \citenamefont {Assumpcao}, \citenamefont {McNally},
  \citenamefont {Burns}, \citenamefont {Sankey},\ and\ \citenamefont
  {Regal}}]{reetz2019analysis}%
  \BibitemOpen
  \bibfield  {author} {\bibinfo {author} {\bibfnamefont {C.}~\bibnamefont
  {Reetz}}, \bibinfo {author} {\bibfnamefont {R.}~\bibnamefont {Fischer}},
  \bibinfo {author} {\bibfnamefont {G.~G.}\ \bibnamefont {Assumpcao}}, \bibinfo
  {author} {\bibfnamefont {D.~P.}\ \bibnamefont {McNally}}, \bibinfo {author}
  {\bibfnamefont {P.~S.}\ \bibnamefont {Burns}}, \bibinfo {author}
  {\bibfnamefont {J.~C.}\ \bibnamefont {Sankey}}, \ and\ \bibinfo {author}
  {\bibfnamefont {C.~A.}\ \bibnamefont {Regal}},\ }\href@noop {} {\bibfield
  {journal} {\bibinfo  {journal} {Physical Review Applied}\ }\textbf {\bibinfo
  {volume} {12}},\ \bibinfo {pages} {044027} (\bibinfo {year}
  {2019})}\BibitemShut {NoStop}%
\bibitem [{\citenamefont {Turner}\ \emph {et~al.}(2001)\citenamefont {Turner},
  \citenamefont {Bock}, \citenamefont {Beeman}, \citenamefont {Glenn},
  \citenamefont {Hargrave}, \citenamefont {Hristov}, \citenamefont {Nguyen},
  \citenamefont {Rahman}, \citenamefont {Sethuraman},\ and\ \citenamefont
  {Woodcraft}}]{turner2001silicon}%
  \BibitemOpen
  \bibfield  {author} {\bibinfo {author} {\bibfnamefont {A.~D.}\ \bibnamefont
  {Turner}}, \bibinfo {author} {\bibfnamefont {J.~J.}\ \bibnamefont {Bock}},
  \bibinfo {author} {\bibfnamefont {J.~W.}\ \bibnamefont {Beeman}}, \bibinfo
  {author} {\bibfnamefont {J.}~\bibnamefont {Glenn}}, \bibinfo {author}
  {\bibfnamefont {P.~C.}\ \bibnamefont {Hargrave}}, \bibinfo {author}
  {\bibfnamefont {V.~V.}\ \bibnamefont {Hristov}}, \bibinfo {author}
  {\bibfnamefont {H.~T.}\ \bibnamefont {Nguyen}}, \bibinfo {author}
  {\bibfnamefont {F.}~\bibnamefont {Rahman}}, \bibinfo {author} {\bibfnamefont
  {S.}~\bibnamefont {Sethuraman}}, \ and\ \bibinfo {author} {\bibfnamefont
  {A.~L.}\ \bibnamefont {Woodcraft}},\ }\href@noop {} {\bibfield  {journal}
  {\bibinfo  {journal} {Applied Optics}\ }\textbf {\bibinfo {volume} {40}},\
  \bibinfo {pages} {4921} (\bibinfo {year} {2001})}\BibitemShut {NoStop}%
\bibitem [{\citenamefont {H{\o}j}\ \emph {et~al.}(2022)\citenamefont {H{\o}j},
  \citenamefont {Hoff},\ and\ \citenamefont {Andersen}}]{hoj2022ultra}%
  \BibitemOpen
  \bibfield  {author} {\bibinfo {author} {\bibfnamefont {D.}~\bibnamefont
  {H{\o}j}}, \bibinfo {author} {\bibfnamefont {U.~B.}\ \bibnamefont {Hoff}}, \
  and\ \bibinfo {author} {\bibfnamefont {U.~L.}\ \bibnamefont {Andersen}},\
  }\href@noop {} {\bibfield  {journal} {\bibinfo  {journal} {arXiv preprint
  arXiv:2207.06703}\ } (\bibinfo {year} {2022})}\BibitemShut {NoStop}%
\bibitem [{\citenamefont {J{\"o}ckel}\ \emph {et~al.}(2011)\citenamefont
  {J{\"o}ckel}, \citenamefont {Rakher}, \citenamefont {Korppi}, \citenamefont
  {Camerer}, \citenamefont {Hunger}, \citenamefont {Mader},\ and\ \citenamefont
  {Treutlein}}]{jockel2011spectroscopy}%
  \BibitemOpen
  \bibfield  {author} {\bibinfo {author} {\bibfnamefont {A.}~\bibnamefont
  {J{\"o}ckel}}, \bibinfo {author} {\bibfnamefont {M.~T.}\ \bibnamefont
  {Rakher}}, \bibinfo {author} {\bibfnamefont {M.}~\bibnamefont {Korppi}},
  \bibinfo {author} {\bibfnamefont {S.}~\bibnamefont {Camerer}}, \bibinfo
  {author} {\bibfnamefont {D.}~\bibnamefont {Hunger}}, \bibinfo {author}
  {\bibfnamefont {M.}~\bibnamefont {Mader}}, \ and\ \bibinfo {author}
  {\bibfnamefont {P.}~\bibnamefont {Treutlein}},\ }\href@noop {} {\bibfield
  {journal} {\bibinfo  {journal} {Applied physics letters}\ }\textbf {\bibinfo
  {volume} {99}},\ \bibinfo {pages} {143109} (\bibinfo {year}
  {2011})}\BibitemShut {NoStop}%
\bibitem [{\citenamefont {St-Gelais}\ \emph {et~al.}(2019)\citenamefont
  {St-Gelais}, \citenamefont {Bernard}, \citenamefont {Reinhardt},\ and\
  \citenamefont {Sankey}}]{st2019swept}%
  \BibitemOpen
  \bibfield  {author} {\bibinfo {author} {\bibfnamefont {R.}~\bibnamefont
  {St-Gelais}}, \bibinfo {author} {\bibfnamefont {S.}~\bibnamefont {Bernard}},
  \bibinfo {author} {\bibfnamefont {C.}~\bibnamefont {Reinhardt}}, \ and\
  \bibinfo {author} {\bibfnamefont {J.~C.}\ \bibnamefont {Sankey}},\
  }\href@noop {} {\bibfield  {journal} {\bibinfo  {journal} {ACS Photonics}\
  }\textbf {\bibinfo {volume} {6}},\ \bibinfo {pages} {525} (\bibinfo {year}
  {2019})}\BibitemShut {NoStop}%
\bibitem [{\citenamefont {Sadeghi}\ \emph {et~al.}(2020)\citenamefont
  {Sadeghi}, \citenamefont {Tanzer}, \citenamefont {Luhmann}, \citenamefont
  {Piller}, \citenamefont {Chien},\ and\ \citenamefont
  {Schmid}}]{sadeghi2020thermal}%
  \BibitemOpen
  \bibfield  {author} {\bibinfo {author} {\bibfnamefont {P.}~\bibnamefont
  {Sadeghi}}, \bibinfo {author} {\bibfnamefont {M.}~\bibnamefont {Tanzer}},
  \bibinfo {author} {\bibfnamefont {N.}~\bibnamefont {Luhmann}}, \bibinfo
  {author} {\bibfnamefont {M.}~\bibnamefont {Piller}}, \bibinfo {author}
  {\bibfnamefont {M.-H.}\ \bibnamefont {Chien}}, \ and\ \bibinfo {author}
  {\bibfnamefont {S.}~\bibnamefont {Schmid}},\ }\href@noop {} {\bibfield
  {journal} {\bibinfo  {journal} {Physical Review Applied}\ }\textbf {\bibinfo
  {volume} {14}},\ \bibinfo {pages} {024068} (\bibinfo {year}
  {2020})}\BibitemShut {NoStop}%
\bibitem [{\citenamefont {Catalini}\ \emph {et~al.}(2020)\citenamefont
  {Catalini}, \citenamefont {Tsaturyan},\ and\ \citenamefont
  {Schliesser}}]{catalini2020soft}%
  \BibitemOpen
  \bibfield  {author} {\bibinfo {author} {\bibfnamefont {L.}~\bibnamefont
  {Catalini}}, \bibinfo {author} {\bibfnamefont {Y.}~\bibnamefont {Tsaturyan}},
  \ and\ \bibinfo {author} {\bibfnamefont {A.}~\bibnamefont {Schliesser}},\
  }\href@noop {} {\bibfield  {journal} {\bibinfo  {journal} {Physical Review
  Applied}\ }\textbf {\bibinfo {volume} {14}},\ \bibinfo {pages} {014041}
  (\bibinfo {year} {2020})}\BibitemShut {NoStop}%
\end{thebibliography}%

\end{document}